\documentclass[9pt,twocolumn,twoside]{pnas-new}
\usepackage{times}
\usepackage{amsfonts}
\usepackage{amssymb}%% The amssymb package provides various useful mathematical symbols
\usepackage{amsmath}
\usepackage{natbib}

\usepackage{graphicx}% Include figure files
\usepackage[tight,footnotesize]{subfigure}
\usepackage{enumitem}

\usepackage{rotating}

\usepackage{xcolor}

\usepackage{setspace}%used to allow caption in table use the whole page size

\templatetype{pnasresearcharticle} % Choose template
\title{Solution to the hyperon puzzle using dark matter}

\author[a,b,c,d,1]{Antonino ~Del Popolo}
\author[e,2]{Maksym Deliyergiyev} 
\author[f,g,3]{Morgan~Le~Delliou}

\affil[a]{Dipartimento di Fisica e Astronomia, University Of Catania, Viale Andrea Doria 6, 95125, Catania, Italy}
\affil[b]{INFN Sezione di Catania, Via S. Sofia 64, I-95123 Catania, Italy}
\affil[c]{Institute of Astronomy, Russian Academy of Sciences, Pyatnitskaya str. 48, 119017 Moscow, Russia}
\affil[d]{Institute of Modern Physics, Chinese Academy of Sciences, Post Office Box 31, Lanzhou 730000, Peoples Republic of China}

\affil[e]{Institute of Physics, Jan Kochanowski University, PL-25406 Kielce, Poland}
\affil[f]{Institute of Theoretical Physics, School of Physical Science and Technology, Lanzhou University, No.222, South Tianshui Road, Lanzhou, Gansu 730000, China}
\affil[g]{Instituto de Astrof\'isica e Ci\^encias do Espa\c co, Universidade de Lisboa, Faculdade de Ci\^encias, Ed. C8, Campo Grande, 1769-016 Lisboa, Portugal}

\leadauthor{A.~Del Popolo}

\correspondingauthor{
\textsuperscript{1} adelpopolo@oact.inaf.it;\\
\textsuperscript{2} maksym.deliyergiyev@outlook.com;\\
\textsuperscript{3} delliou@lzu.edu.cn; delliou@ift.unesp.br
}


\begin{abstract}
In this paper, we studied the ``hyperon puzzle", a problem that nevertheless the large number of studies is still an open problem. The solution of this issue requires one or more mechanisms that could eventually provide the additional repulsion needed to make the EoS stiffer and, therefore, the value of $M_{\rm{max}, T}$ compatible with the current observational limits. In this paper we proposed that including dark matter (DM) admixed with ordinary matter in neutron stars (NSs), change the hydrostatic equilibrium and may explain the observed discrepancies, regardless to hyperon multi-body interactions, which seem to be unavoidable. 

We have studied how non-self-annihilating, and self-interacting, DM admixed with ordinary matter in NSs changes their inner structure, and discussed the mass-radius relations of such NSs. We considered DM particle masses of 1, 10 and 100 GeV, while taking into account a rich list of the DM interacting strengths, $y$. 

By analyzing the multidimensional parameter space, including several quantities like: a. the DM interacting strength, b. the DM particle mass as well as the quantity of DM in its interior, and c. the DM fraction, ${\rm f}_{DM}$, we put constraints in the parameter space ${\rm f}_{DM} - p^{\prime}_{\rm DM}/p^{\prime}_{\rm OM}$. Our bounds are sensitive to the recently observed NSs total masses. 
\end{abstract}

\doi{Submitted to Physics of the Dark Universe}

\begin{document}
\maketitle
\thispagestyle{firststyle}
\ifthenelse{
	       \boolean{shortarticle}
           }
{\ifthenelse{
		   \boolean{singlecolumn}
	       }
	{\abscontentformatted}{\abscontent}
}{}	

%%%%%%%%%%%%%%%%%%%%%%%%%%%%%%%%%%%%%%%%%%%%%%%%%%%%%%
%%%%%%%%%%%%%%%%%%%%%%%%%%%%%%%%%%%%%%%%%%%%%%%%%%%%%%
\section{Introduction}

As the extreme conditions reached in Neutron Stars (hereafter NSs) cannot be reached on Earth's laboratories, their natural environment can be the only one permitting investigations of the fundamental constituents of matter and their interactions under such extreme conditions. Since their internal constituents and bulk properties are the result of the strong interaction of their matter \citep{Prakash:1996xs, Lattimer:2004pg}, they probe the Equation of State (EoS) of matter in such extreme conditions, i.e. the thermodynamical constrain between the matter's pressure, energy density, and temperature.
Data collected from ground-based radio telescopes and several generations of $X$-ray and $\gamma$-ray satellites on NSs has fueled a major effort of the last few decades, directed at solving the EoS model properly describing those NSs, a fundamental problem of  astrophysics and nuclear physics.
New pathways in the study of NS matter physics have recently opened with gravitational waves observation from the GW170817 NS-NS merger event \citep{TheLIGOScientific:2017qsa} and its electromagnetic transient counterpart \citep[AT 2017gfo, Ref.][]{GBM:2017lvd}, that allowed to identify NS merger as an important source for $r$-process nuclei \citep{Ellis:2017jgp, Ellis:2018bkr, Tews:2019cap, Schuetrumpf:2019mcf}. This observation additionally yielded constraints on mass and tidal deformability \citep{Most:2018hfd}.
However, it should be noted that some authors \citep{Zhang:2018vrx, Lim:2018bkq, Tews:2019cap} claimed this combined observation doesn't add new insight on the NS EoS,  
possibly conceeding constraints for the stellar radius lower limit \citep{Bauswein:2017vtn, Radice:2017lry, Radice:2018ozg, Koppel:2019pys}, if systematic error effects are set aside \citep{Tews:2019cap}.

In the instance when the NS is only considered as a Fermi neutrons gas, the balance between the gravitational attraction and the degenerate neutron gas pressure determines its equilibrium structure. Its mass is thus obtained from the Oppenheimer-Volkoff limit ($\simeq 0.7 M_{\odot}$). When strong interactions are considered in its structure, the NS mass increases from the resulting repulsive force between neutrons. 
Such a model is however incomplete, as several studies \citep{Bombaci:2016xzl} have shown the emergence of hyperons (also noted $Y$s) to be unavoidable at $2-3$ times the usual nuclear saturation density  \citep[$\rho_{0}=0.16$ fm$^{-3}\approx 2.7\cdot 10^{14}$ $\rm{g/cm}^{3}$, see][]{Balberg:1998ug}, at which nuclei completely dissolve, and the NS core begins. For such density, conversion of nucleons into hyperons becomes energetically favorable, as neutron and proton chemical potentials reach large enough levels. 
The remarkable effect of this conversion, that much softens the EoS to the point that it releases off the Fermi pressure by the baryons, is to cancel the strong three-nucleon repulsion (TNR) impact on the maximum mass \citep{Baldo:1999rq,Vidana:2000ew,Nishizaki:2001in}, resulting in  the model's theoretically predicted maximum NS mass decrease from $M_{\rm{max}}=2.28 M_{\odot}$ to $M_{\rm{max}}=1.38 M_{\odot}$ \citep{Bombaci:2016giv}. Such $M_{\rm{max}} < 2 M_{\odot}$ prediction can be found in many hyperon stars structure calculations, especially the microscopic hyperonic matter EoSs based types \citep{Li:2008zzt, Vidana:2000ew, Schulze:2011zza, Djapo:2008au}.
In addition to this softening, the $Y$-mixing generates very efficient $\nu$ - emission processes (i.e., $\beta$-decay in the presence of $Y$)\footnote{So-called ``hyperon direct URCA ($Y$-DUrca)'', yielding $(5\sim 6)$ orders of magnitude larger emissivity than the standard modified URCA (MUrca) processes, described by $n+N\rightarrow p+N+e^{-}+\bar{\nu}_{e}$ and $p+N+e^{-}\rightarrow n+N+\nu_{e}$.} that should accelerate the cooling of NSs inconsistently with recent surface-temperature observations, and thus require some suppression mechanism \citep{Tsuruta_2009}. The first or both problems are dubbed ``hyperon puzzle'' or ``hyperon crysis'' of neutron stars \citep{Chatterjee:2015pua, Bombaci:2016xzl}.

Apart from hyperons, several models, such as free quarks \citep{Alford:2007xm}, and mesons predict non-nucleonic components in NS interiors, with similar softening effects on the EoS, and its subsequent NS theoretical maximum mass reduction \citep{Burgio:2010ek, Burgio:2011wt, Zuo:2004az, Li:2006qn, Li:2008ab, Peng:2008ta, Li:2010yc, LiAng:4233, Ang_2009}, that can render such mass incompatible with the current largest NS masses measured at $~2M_{\odot}$ \citep{Demorest:2010bx,Arzoumanian:2017puf, Cromartie:2019kug}. 
Such $2 M_{\odot}$ NSs cannot, therefore, be obtained with a Thomas-Fermi model for non-uniform matter \citep{Shen:1998gq} including hyperons \citep{Ishizuka:2008gr, Shen:2011qu} or a quark-hadron phase transition \citep{Nakazato:2008su}. Furthermore, EoS models unable to support high values for the stellar masses have been ruled out by observations: $1.8M_{\odot}$ for Vela X-1\citep{Rawls:2011jw}, $2M_{\odot}$ for 4U 1822-371 \citep{MunozDarias:2005jz},
PSR J1614-2230 with $M=1.97 \pm 0.04 M_\odot$ \citep{Demorest:2010bx} and PSR J0348+0432 of $M = 2.01 \pm 0.04M_\odot$ \citep{Antoniadis:2013pzd}\footnote{Recall that $3.2 M_\odot$ is the NSs mass upper limit from General Relativity, while gravitational waves observations give NSs  masses $\lesssim 2.2 M_\odot$ \citep{Margalit:2017dij, Rezzolla2017}. We further have a NSs mass theoretical lower limit at 0.1 $M_{\odot}$, however, below about 1 $M_{\odot}$, lepton-rich proto NSs are unbound \citep{Lattimer:2004pg}.}.

Several proposals attempt 
to solve the ``hyperon puzzle''. Stiff EoS, reconciling theory with the observed NS mass of $2M_{\odot}$, suggests the existence of strongly repulsive many-body effects (MBE) in the high-density region, and strongly interacting quark matter (i.e., hybrid star) troubleshoots the problem \citep{Burgio:2002sn, Schulze:2011zza}. 
Strong repulsions are needed not only in nucleon-nucleon-nucleon 
($NNN$) but also in hyperon-nucleon-nucleon ($YNN$) and hyperon-hyperon-nucleon ($YYN$) channels \citep{Nishizaki:2001in}. However, the literature only presents a few fully consistent calculations including hyperonic three-body forces (YTBF) 
\citep{Vidana:2010ip, Nishizaki:2002ih}, and 
it is moreover difficult to prove the existence of MBE including strong repulsion on the basis of the experimental data of $B_{\Lambda}$, because the two-body interaction model is not well determined. 
NSs with masses larger than 2 $M_{\odot}$ seem to be out of reach even for hybrid stars. Hyperonic three-body interactions do not seem to provide the full solution of the `hyperon puzzle', although they probably contribute to its solution \citep{Logoteta:2019utx}.

The pure neutron matter EoS gets strong  constraints from calculations using the Auxiliar Field Diffusion Monte Carlo method \citep{Gandolfi:2019zpj}. However, as NS does not only reduce to neutron matter, those calculations need to be extended to more realistic NS compositions. The following will discuss inclusion of dark matter (DM) into the extension 
NSs containing DM solves many problems. 
If ambient DM particles scatter with ordinary matter in stars, they can lose kinetic energy and become gravitationally bound by the star \citep{Gould:1987ju,
	Gould:1987ir,
	Jungman:1995df,
	Kappl:2011kz,
	Busoni:2013kaa,
	Bramante:2017xlb}. 
This should modify the local pressure-energy density relationship of the matter and hence change the theoretical prediction of the gravitational mass of the star. 
Admixing DM with NS matter, coupled only through gravity, produces similar results to that of exotic states \citep{Sandin2009,Ciarcelluti:2010ji,Leung:2011zz,Li:2012ii,Xiang:2013xwa,Ellis:2018bkr}. As shown by \citep{Ciarcelluti:2010ji}, DM allows to explain very compact NSs \citep{Ciarcelluti:2010ji}, or very massive pulsars ($2M_{\odot}$, e.g., PSR J1614-2230 with $M=1.97 \pm 0.04 M_\odot$ \citep{Demorest:2010bx} and PSR J0348+0432 of $M = 2.01 \pm 0.04M_\odot$ \citep{Antoniadis:2013pzd}). The NS can get to $2M_{\odot}$ for DM ratio of 15\% according to Ref.\citep[and masses  of $1.8 M_{\odot}$ for a DM ratio of 70\%]{Ciarcelluti:2010ji} 
or for DM ratio of 50\% according to \citep{Goldman:2013qla}. The precise NS mass depends also on the DM particle mass \citep{Li:2012ii,Mukhopadhyay:2015xhs}, so the final NS mass results from a combination of relative acquired DM mass and DM particle mass \citep{Leung:2011zz}.
Several authors studied admixed NSs, e.g., with mirror matter \citep{Sandin2009}, degenerate DM \citep{Leung2011}, and ADM \citep{Li:2012ii} finding that increasing the ratio of DM to normal matter yielded stars with smaller radii and masses. 
Ref.~\citep{Ellis:2017jgp} studied the imprints of the possible presence of DM cores with the NSs in the power spectral density of the gravitational wave emission following a NS-NS merger. 
Ref.~\citep{Ellis:2018bkr} studied the changes of tidal deformability parameter, $\Lambda$,\footnote{Not to be mistaken for 
	the $\bar{\Lambda}$ used later in the paper to denote the minimal subtraction scheme scale.} for DM admixed NS, and compared its LIGO/Virgo upper limit to constrain the EoS.
Ref.~\citep{Mukhopadhyay:2015xhs} studied quark matter admixed DM, finding a star mass 1.95 $M_{\odot}$, while \citep{Li:2012ii, LI:2013qqa} obtained NSs masses $\simeq 2 M_{\odot}$ for $m_{\rm dm} \simeq 0.1$ GeV (for weakly interacting DM), or $\simeq 1$ GeV (for strongly interacting DM). Ref.~\citep{Tolos2015} studied NSs, and WDs matter admixed with 100 GeV ADM, finding that planets-like objects could form. Ref.~\citep{Tolos2015}'s study was extended in \citep{Deliyergiyev:2019vti} to 
particle masses in the range 1-500 GeV, finding, among other results, an increase of the compact objects (COs) mass with decreasing acquired DM particles masses, and that the smaller the DM particle mass, the more DM is captured in the COs.
Ref.~\citep{Ding:2019cky} found that DM affects the NS cooling rate. Compared with the ordinary matter NS, the light DM particles, $0.2 \leq m_{\rm dm} \leq  0.4$ GeV, favor the NS cooling and make the massive NS cool faster; 
for intermediate DM mass particles, $0.6 \leq m_{\rm dm} \leq  0.8$ GeV, the cooling of the NS is slowed down compared with DM free NSs; and for larger DM mass particles, $0.8 \leq m_{\rm dm} \leq  1.2$ GeV, the Urca process can appear in very low mass NSs, $\sim 1 M_{\odot}$, contrary to normal NSs.

As one can study the effects of DM on the NSs EoS, and the change of their structure, conversely 
NSs can be used as a probe of the existence and nature of DM. 
When the DM agglomeration exceeds a critical mass, the DM forms a black hole (BH) at the NS center that may consume the NS \citep{Kouvaris:2013awa}\footnote{NS collapses into the BH on the dynamical time scale ($t_{\rm{dyn}}\sim 1$ ms), and a  $\sim1.4 M_{\odot}$ BH remnant is left behind \citep{Fuller:2014rza}.}, allowing to pose constraints on the DM nature \citep{Kouvaris:2013awa}.
In two recent papers \citep{Kouvaris:2010vv, deLavallaz:2010wp}, constraints on the properties of DM, in the form of weakly interacting massive particles (WIMPs), were put by considering the effects on the luminosity of the accretion and self-annihilation of DM in NS.
In particular, WIMPs annihilation in DM cores should produce temperature and luminosity changes, through heat, of old stars \citep{Kouvaris2008, Bertone2008, Kouvaris:2010vv, deLavallaz:2010wp}. 
However, those changes \citep{Kouvaris2008} are below observed temperatures \citep{Sandin2009} and are difficult to observe \citep{Kouvaris2008,Sandin2009}.
Different is the situation close to the galactic center (GC), where NSs have a larger surface temperature, $10^6$ K, and luminosities $10^{-2} L_{\odot}$ \citep{deLavallaz:2010wp}. However, the detection of those temperature changes is also very difficult in this case, 
due to (e.g.) dust absorption
\citep{deLavallaz:2010wp}.
This implies that WIMPS effects on NSs are of little practical concern, while other forms of DM can have testable effects (e.g. asymmetric dark matter, i.e. ADM). In this work, we will focus on the non-self-annihilating, and self-interacting, particle, such as the newly interesting mirror DM \citep[and references therein]{Okun:2006eb} or asymmetric DM \citep[and references therein]{Goldman:2013qla}. In particular, their accumulation on NSs results in a rather sensitive maximum mass to the EoS model of this ADM.

A recently discovered class of radio transients, such as fast radio bursts \citep{Keane:2012yh, Thornton:2013iua, Burke-Spolaor:2014rqa, Spitler:2014fla}, provides an intriguing possibility in the detection of 
NS with sufficiently  accreted 
DM near the GC \citep{Falcke:2013xpa, Fuller:2014rza}. 
The collapse of the DM core may trigger the further collapse of the NS into a BH with emitting fast radio bursts, and disrupt the host star.
The accreted DM scales inversely proportional to the NSs life time close to GC, as $t_{c}\propto \rho_{DM}^{-1}$, since the DM capture rate is proportional to the ambient DM density, $\rho_{DM}$.
The complementary probe for such models could be an observation of the NS mass change induced by agglomerated DM with respect to the galactocentric distances proposed by \citep{DelPopolo:2019nng}. In other words, addressing the problem of exotic relativistic stars in GR, could also represent a testbed for DM. 

We show that it is possible to address simultaneously the maximal value of NS mass as well as fit the data by \citep{Ozel:2010fw} assuming DM contribution to the NS material.
The paper is organized as follows. In Sec.(\ref{sec:TOVeq_admixDM}) we briefly recapitulate the structure equation for compact stars, the TOV equation, whereas in Sec.(\ref{sec:Results}) we describe the results of our numerical solution of the TOV equations. We show how the critical mass of the core is modified when DM content is present. The mass-radius relation changes such that 
the changes in the radius have latency 
with respect to the total mass, while DM fraction is increasing. We compare our results with the latest NS observations. In Sec. (\ref{sec:Constraints}) we aimed to provide several constraint regions consistent with the currently observed mass range of NSs. Conclusion remarks are given in Sec. (\ref{sec:Conclusions}).

%=====================================
\begin{figure*}%[h]
	\centering
	\includegraphics[scale=0.19]{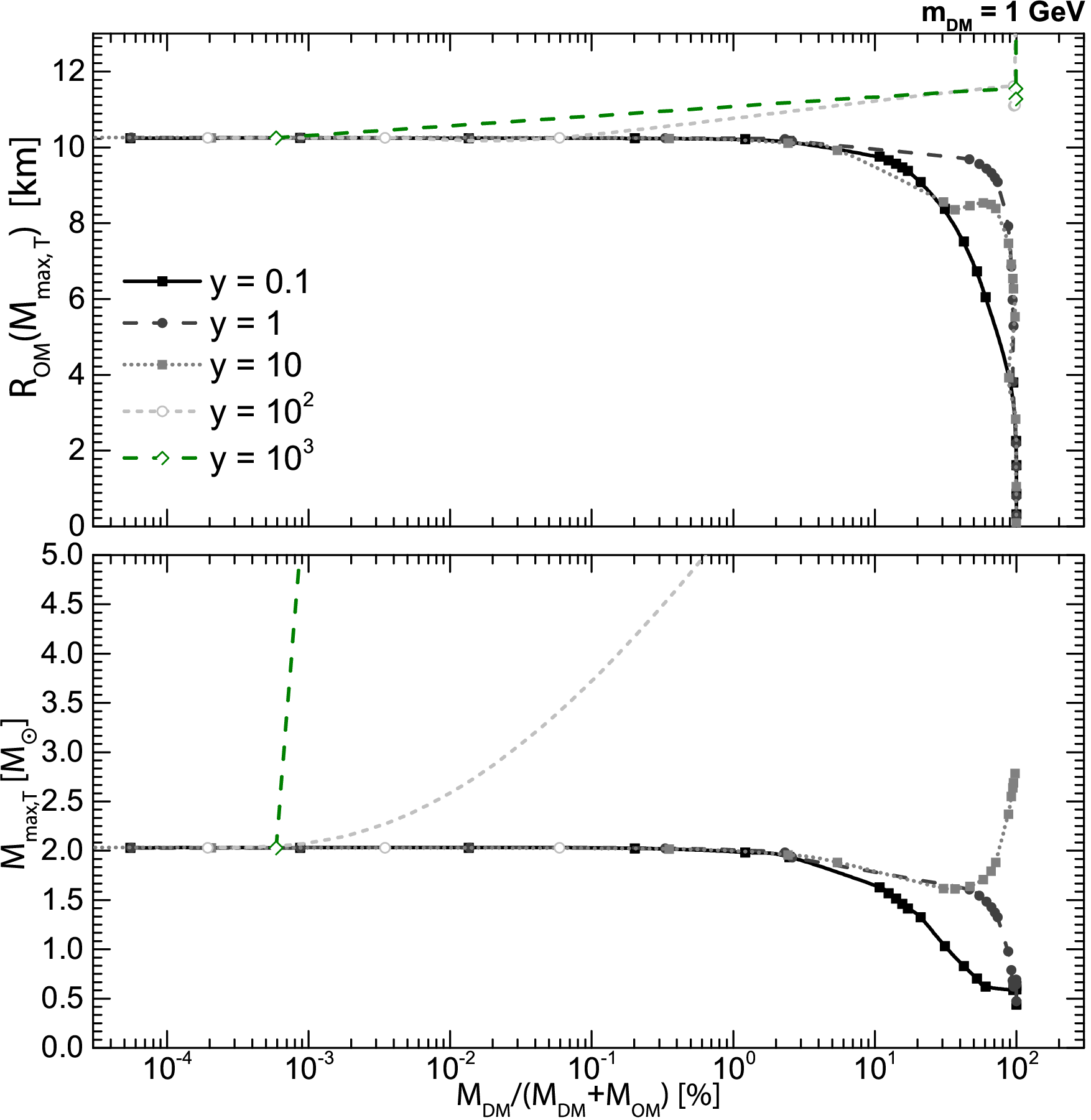%
	}
	\includegraphics[scale=0.19]{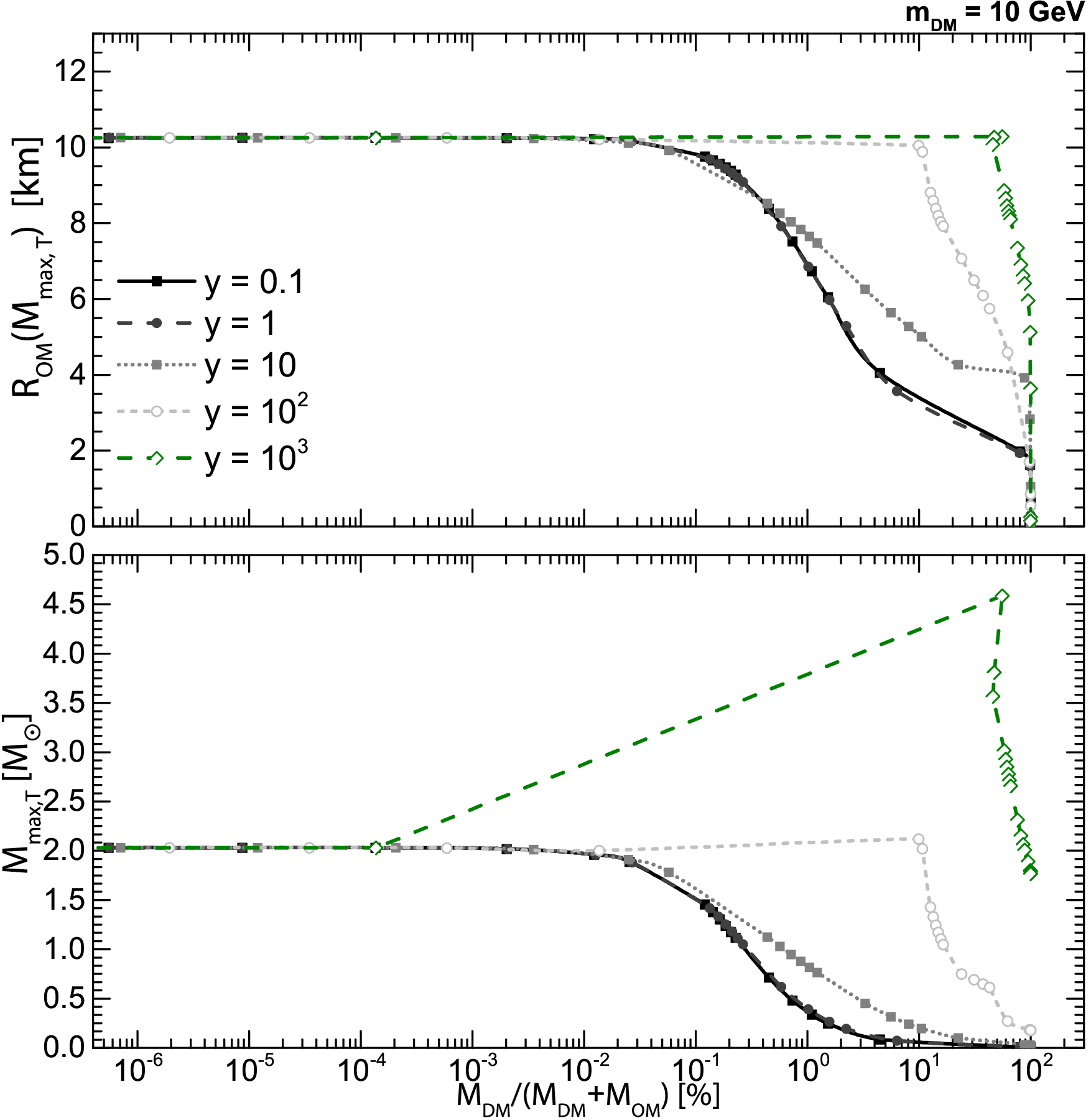%
	}
	\includegraphics[scale=0.19]{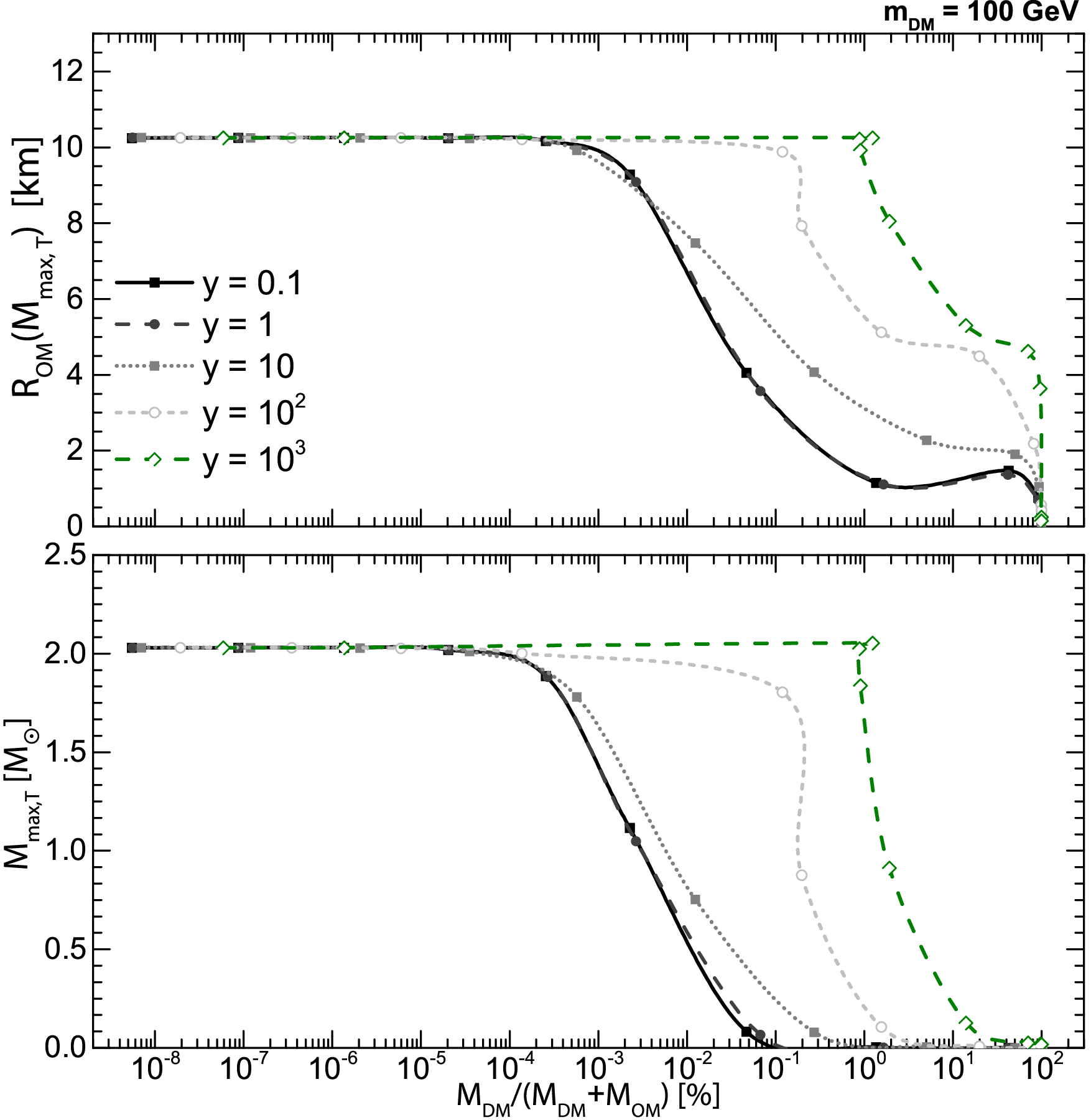%
	}
	\caption{
		(Upper panels) Radii of the stars at the maximum value of the total mass, $M_{T}$, for different ratios of the dimensionless DM pressure ($p^\prime_{DM}$) versus the dimensionless OM pressure ($p^\prime_{OM}$), i.e. $p^\prime_{DM}/p^\prime_{OM}$ (denoted as data points), and maximum total masses (lower panels) vs. the relative number of dark matter particle to ordinary baryons (ordinary matter). The ratios, $p^\prime_{DM}/p^\prime_{OM}$, are increasing from left to right from $10^{-5}$ to $10^{5}$. Results are shown assuming different mass of the hypothetical DM particle in the range from 1 to 100 GeV (from left to right).
	}
	\label{fig:NS_MTvsRom}
\end{figure*}
%=====================================

%=====================================
\begin{figure}%[h]
	\centering
	\includegraphics[scale=0.35]{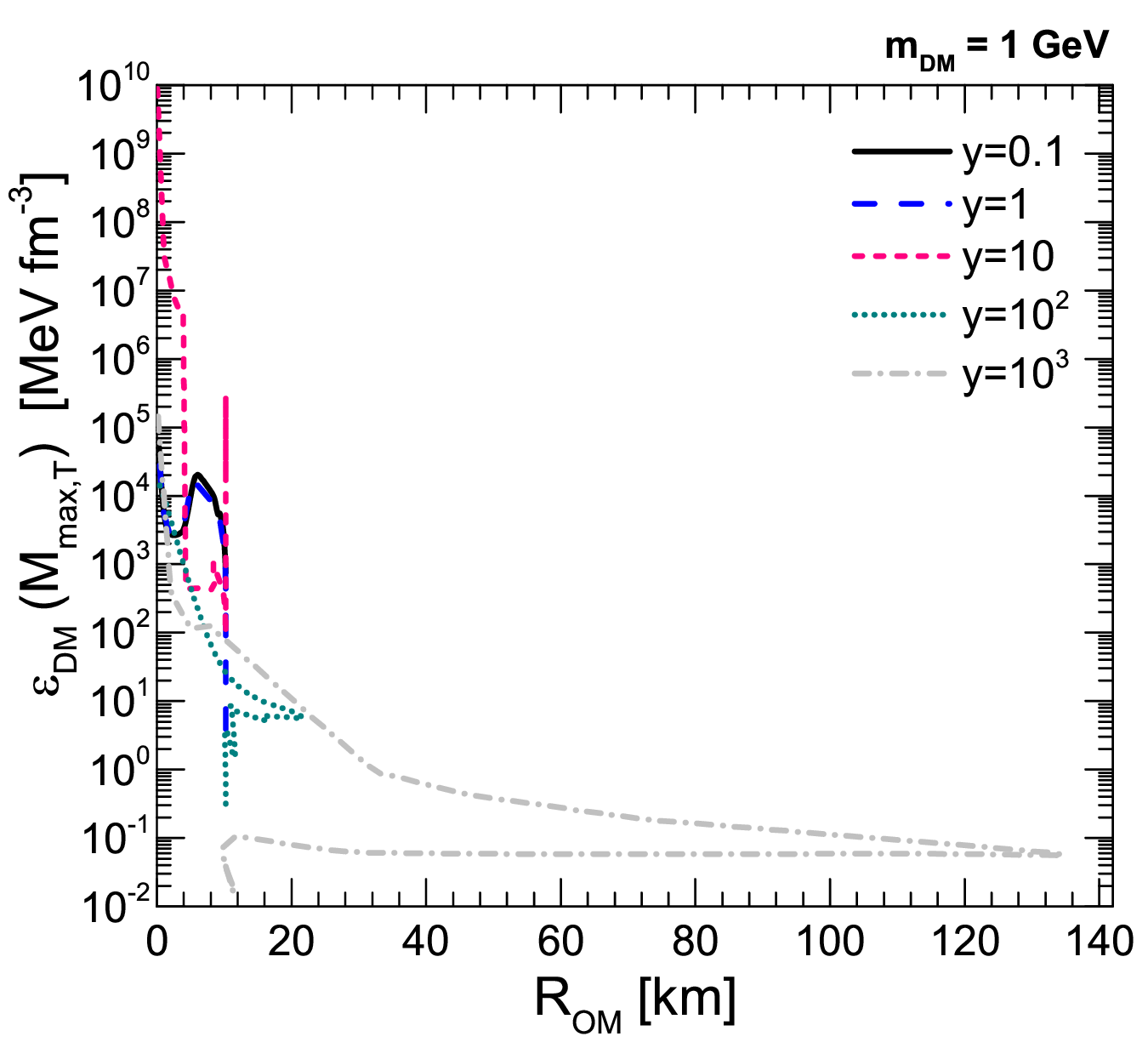}
	\includegraphics[scale=0.35]{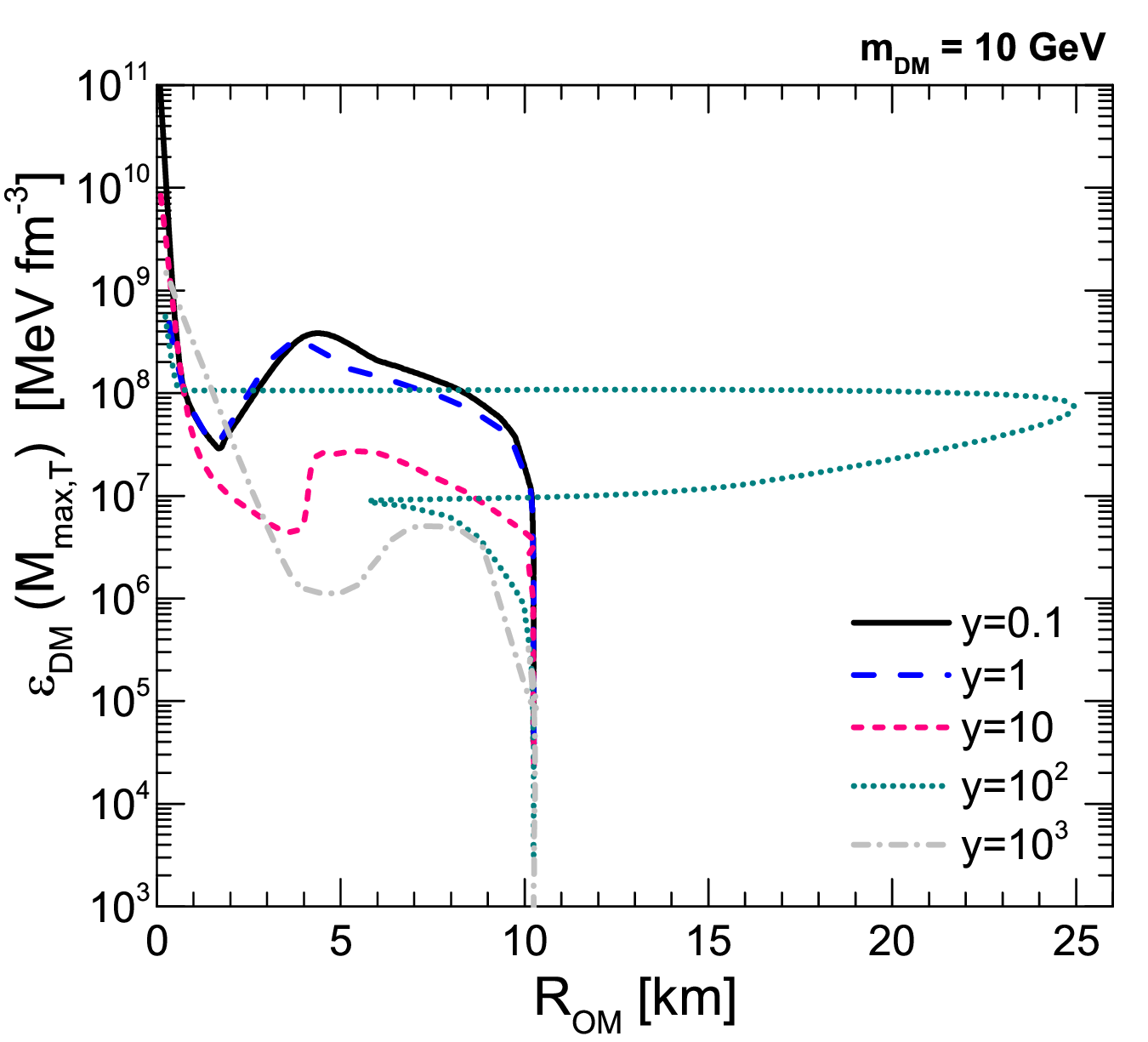}
	\includegraphics[scale=0.35]{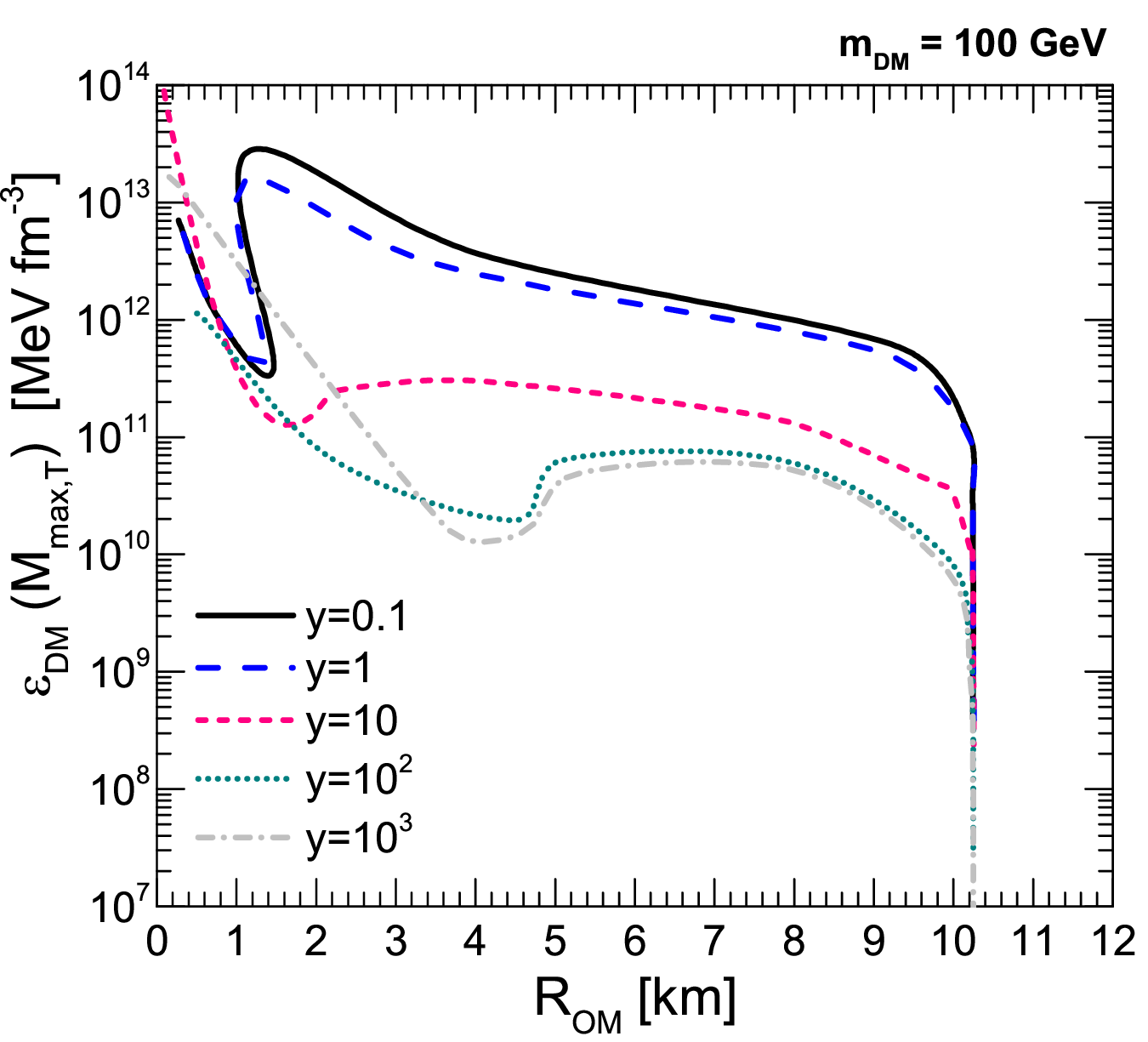}
	\caption{
		Energy density profiles of DM inside NS at the maximal total mass $M_{\rm{max,T}}$, for the range of DM interaction streght parameters, $y=0.1 - 1000$.  Calculations have been performed for $m_{\rm{dm}} =1, 10, 100$ GeV. For the given, $y$, we scan the range of $p^\prime_{DM}/p^\prime_{OM}$, from $10^{-5}$ to $10^{5}$.
	}
	\label{fig:EnerDenstDM_vs_Rom}
\end{figure}
%=====================================

%%%%%%%%%%%%%%%%%%%%%%%%%%%%%%%%%%%%%%%%%%%%%%%%%%%%%%
%%%%%%%%%%%%%%%%%%%%%%%%%%%%%%%%%%%%%%%%%%%%%%%%%%%%%%
\section{Model: Tolman-Oppenheimer-Volkoff Equations with Admixtured Dark Matter}
\label{sec:TOVeq_admixDM} 

In order to study the structure of NSs made of an admixture of ADM and ordinary matter (OM), coupled only by gravity, we have to solve the Tolman-Oppenheimer-Volkoff \citep[hereafter TOV]{Tolman:1939jz,oppenheimer-1939a} equations with physically motivated EoS. 
The dimensionless TOV equations \citep{Narain:2006kx} are given by
%----------------------------------------------------
\begin{align}
	\frac{dp'_{OM}}{dr}=&-(p'_{OM}+\rho'_{OM})\frac{d \nu}{dr}, \nonumber \\
	\frac{dM'_{OM}}{dr}=&4 \pi r^2 \rho'_{OM}, \nonumber \\
	\frac{dp'_{DM}}{dr}=&-(p'_{DM}+\rho'_{DM}) \frac{d \nu}{dr}, \nonumber \\
	\frac{dM'_{DM}}{dr}=&4 \pi r^2 \rho'_{DM}, \nonumber \\
	\frac{d \nu}{dr}=&\frac{(M'_{OM}+M'_{DM}) + 4 \pi r^3(p'_{OM}+p'_{DM})}{r(r-2(M'_{OM}+M'_{DM}))}, 
	\label{eq:TOV_with_DM}
\end{align}
%----------------------------------------------------
where $p'=P/m_f^4$ is the dimensionless pressure, and $\rho'=\rho/m_f^4$ the energy density, being 
$m_f\equiv m_{\rm {dm}}$ the fermion mass (i.e., DM particle mass, and neutron mass). The DM particles are non-self annihilating \citep{Nussinov:1985xr,Kaplan:1991ah,Hooper:2004dc,Kribs:2009fy,Kouvaris:2015rea}, and self-interacting fermions \citep{Spergel:1999mh}. 
Similarly to \citep{Deliyergiyev:2019vti, DelPopolo:2019nng}, the DM particles have masses: 1, 5, 10, 50, 100, 200, or 500 GeV. 
$\frac{d \nu}{dr}$ is the full source for TOV. Note that the TOV equations give rise to DM pressure, even when it is negligible away from the compact object. 
Each one of the two species can give rise to an astrophysical object with radius, $R=(M_p/m_f^2) \, r$ and mass $M= (M_p^3/m_f^2) \,  m$, where $M_p$ is the Planck mass \citep{Narain:2006kx}.

OM has an EoS given by the fitting function to the complicated numerical EoS derived in Ref.\citep{Kurkela:2009gj}, that involves an interpolation between the regimes of low-energy chiral effective theory and high-density perturbative QCD:
\begin{equation} 
	P_{QCD}=P_{SB}(\mu_{B})\left( c_{1} - \frac{a(X)}{(\mu_{B}/{\rm{GeV}})- b(X)} \right)
	\label{eq:OM_EoS}
\end{equation}
where $P_{SB}(\mu_{B})=\frac{3}{4\pi^{2}} (\frac{\mu_{B}}{3})^{4}$ is the Stefan-Boltzmann pressure of three massless non-interacting quark flavors. The functions $a(X)=d_{1}X^{-\nu_{1}}$ and $b(X)=d_{2}X^{-\nu_{2}}$, help to couple results to the renormalization scale, and  depend on a dimensionless parameter proportional to the scale parameter, $X\equiv 3\bar{\Lambda}/\mu_{B}$. Here, $\mu_B$ is the baryon chemical potential, $\bar{\Lambda}$ is the minimal subtraction scheme scale. 
This interpolation shrinks the EoS band to a well defined region in the mass-radius diagram of compact stars \citep[see, e.g.][]{Kurkela:2014vha}. Moreover, we map the EoS with an inner and outer crust EoS using \citep{Negele:1971vb} and \citep{Ruester:2005fm}, respectively. For central density $\rho < 3.3 \times 10^{3}~{\rm{g/cm^{3}}}$ we use the Harrison-Wheeler EoS \citep{Harrison:1965du, Thorne1671}, more details in \ref{sec:Appendix01}.

The \eqref{eq:OM_EoS} was taken from a "global fit", which was obtained by fixing the strong coupling constant and the strange quark mass at arbitrary reference scales (using lattice and experimental data). Hyperons physics is present there through strange quarks interactions, although not explicitly \citep{Kurkela:2009gj}.

%=====================================
\begin{figure}%[h]
	\centering
	\includegraphics[scale=0.3]%{MTmX_pDMpOM103_Ylog_Xlin}%
	{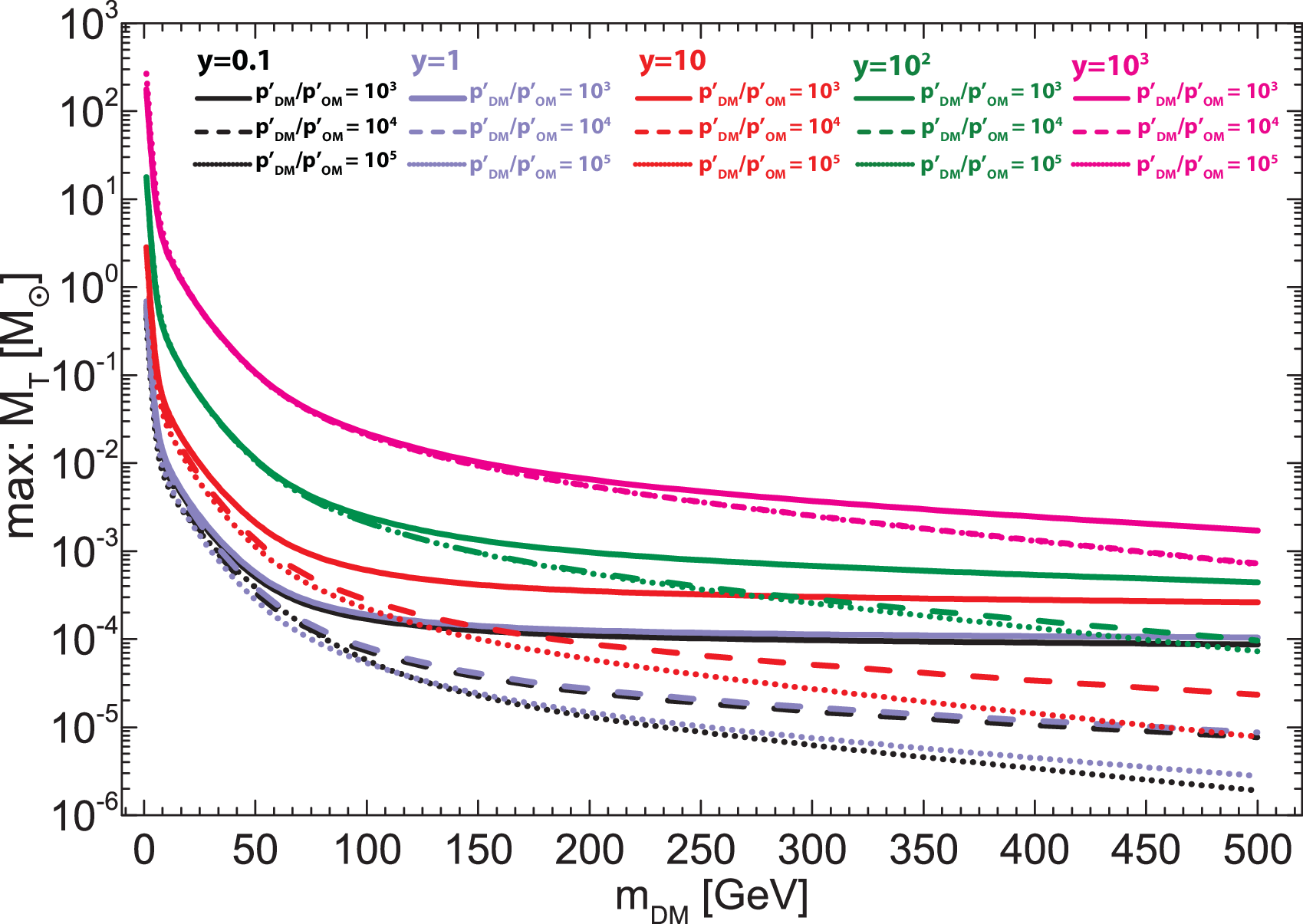}
	%\hspace*{-2.9cm}\begin{minipage}{\dimexpr\textwidth+5.8cm}
		%\subfigure[$p_{\rm{DM}}/p_{\rm{OM}}=10^{3}$]{
			%\includegraphics[width=.45\columnwidth%scale=0.25
			%]{MTmX_pDMpOM103_Ylog_Xlin}
			%}
		%\subfigure[$p_{\rm{DM}}/p_{\rm{OM}}=10^{4}$]{
			%\includegraphics[width=.45\columnwidth%scale=0.25
			%]{MTmX_pDMpOM104_Ylog_Xlin}
			%}
		%\subfigure[$p_{\rm{DM}}/p_{\rm{OM}}=10^{5}$]{
			%\includegraphics[width=.45\columnwidth%scale=0.25
			%]{MTmX_pDMpOM105_Ylog_Xlin}
			%}		
		%\end{minipage}\hspace*{-2.9cm}
		\caption{
			Maximum masses of the stars with the admixtured DM as a function of the particle mass of DM candidates, $m_{\rm{dm}}$. 
			The solid line corresponds to $p^\prime_{\rm{DM}}/p^\prime_{\rm{OM}}=10^{3}$, dashed line is $p^\prime_{\rm{DM}}/p^\prime_{\rm{OM}}=10^{4}$,   dotted line is $p^\prime_{\rm{DM}}/p^\prime_{\rm{OM}}=10^{5}$. By black lines we denote $y=0.1$, blue lines are $y=1$,
			red lines are $y=10$, green lines are $y=10^{2}$, and magenta lines are $y=10^{3}$. 
			Where $y$ is the DM interaction strength parameter \citep{Tolos:2015qra, Deliyergiyev:2019vti}.
		}
		\label{fig:MT_pDMpOM}
	\end{figure}
	%=====================================
	%=====================================
	\begin{figure*}%[h]
		\centering
		%\vspace{-2cm}
		%\begin{minipage}{\dimexpr\textwidth+4cm}
		%\includegraphics[width=1.25\columnwidth]{ALL_mDM_LEGEND}
		%\hspace*{-1.2cm}%-1.6cm}
	\includegraphics[scale=0.55]{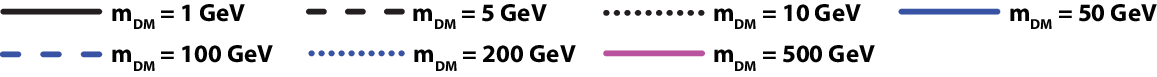}
	
	\includegraphics[scale=0.23]{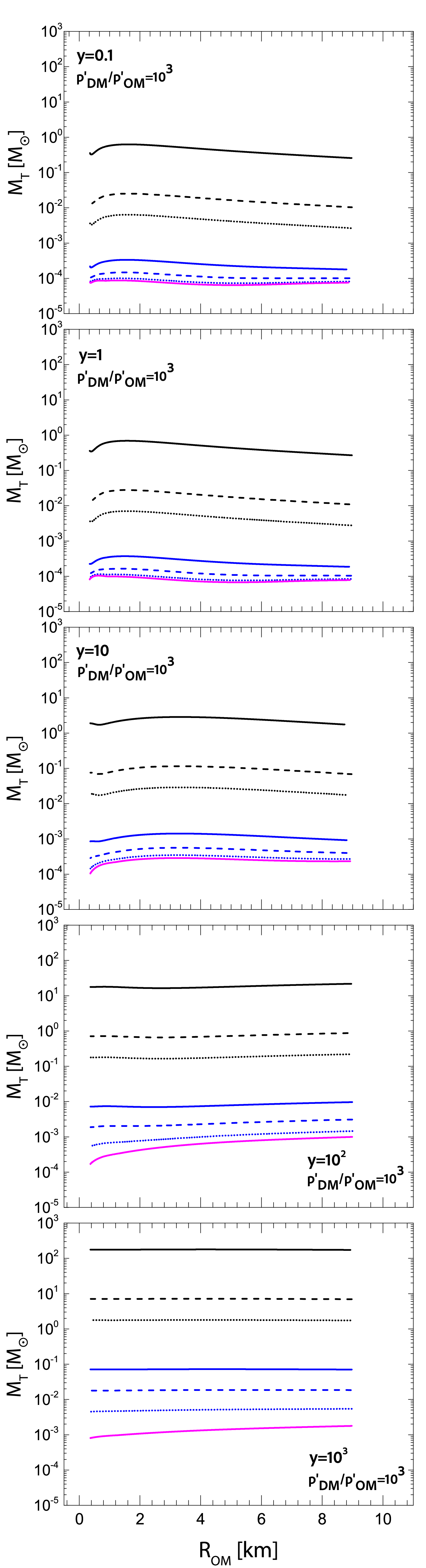}
	\includegraphics[scale=0.23]{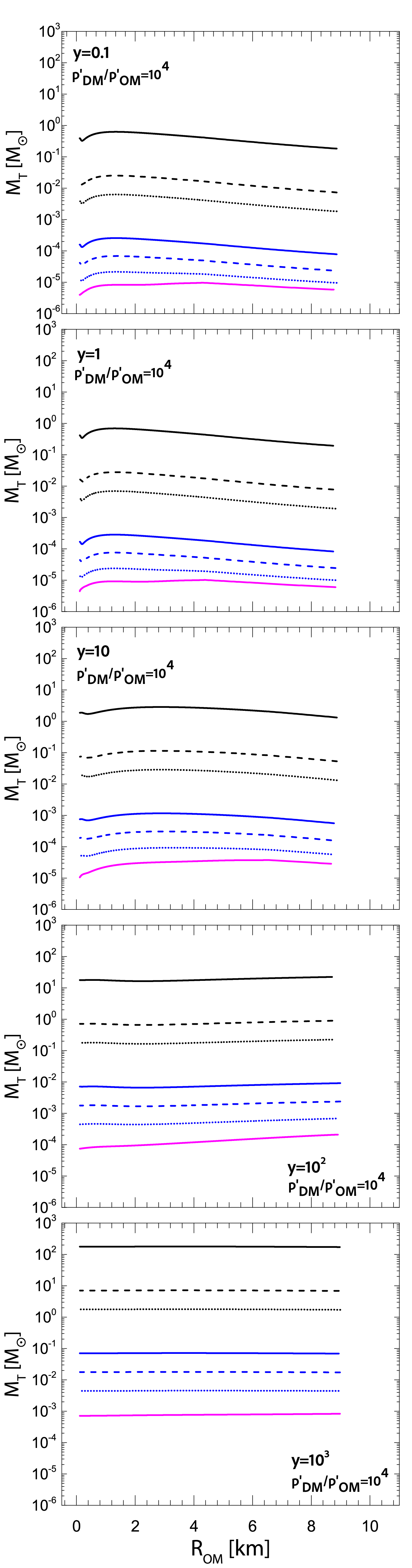}
	\includegraphics[scale=0.23]{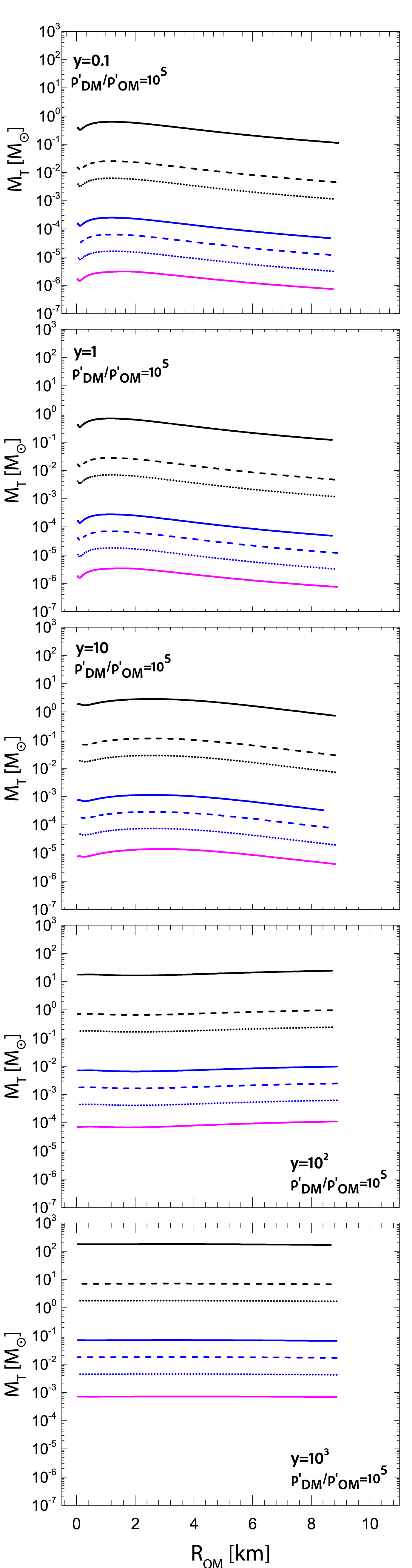}
	%\end{minipage}\hspace*{-2cm}
	\caption{
		Mass-radius relations of the equilibrium configuration of DM-admixed NS branch for the DM interaction strength, $y=0.1; 1; 10; 100; 10^{3}$ (from top to bottom). Results are shown for DM particle mass $m_{\rm dm}$ ranging from 1 to 500 GeV (1, 5, 10, 50, 100, 200, 500). 
		Each column corresponds to a different ratio between the DM pressure and that of the ordinary matter, $p^\prime_{DM}/p^\prime_{OM}$, which is assumed to be in the range of $10^{3}$ to $10^{5}$.
	}
	\label{fig:MTRom_m1_500_y01_103}
\end{figure*}
%=====================================

%=====================================
\begin{figure*}%[h]
\centering
%\vspace{-.5cm}\hspace*{-0.8cm}
%\includegraphics[scale=0.36]{ALL_LEGEND_mod_labels}
\includegraphics[scale=0.32]{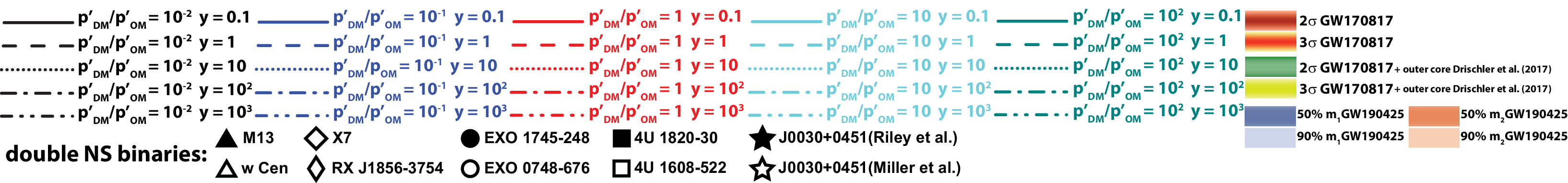}
%\hspace*{-1.8cm}\vspace{-0.5cm}\begin{minipage}{\dimexpr\textwidth+3.6cm}
\subfigure[]{\includegraphics[scale=0.20]{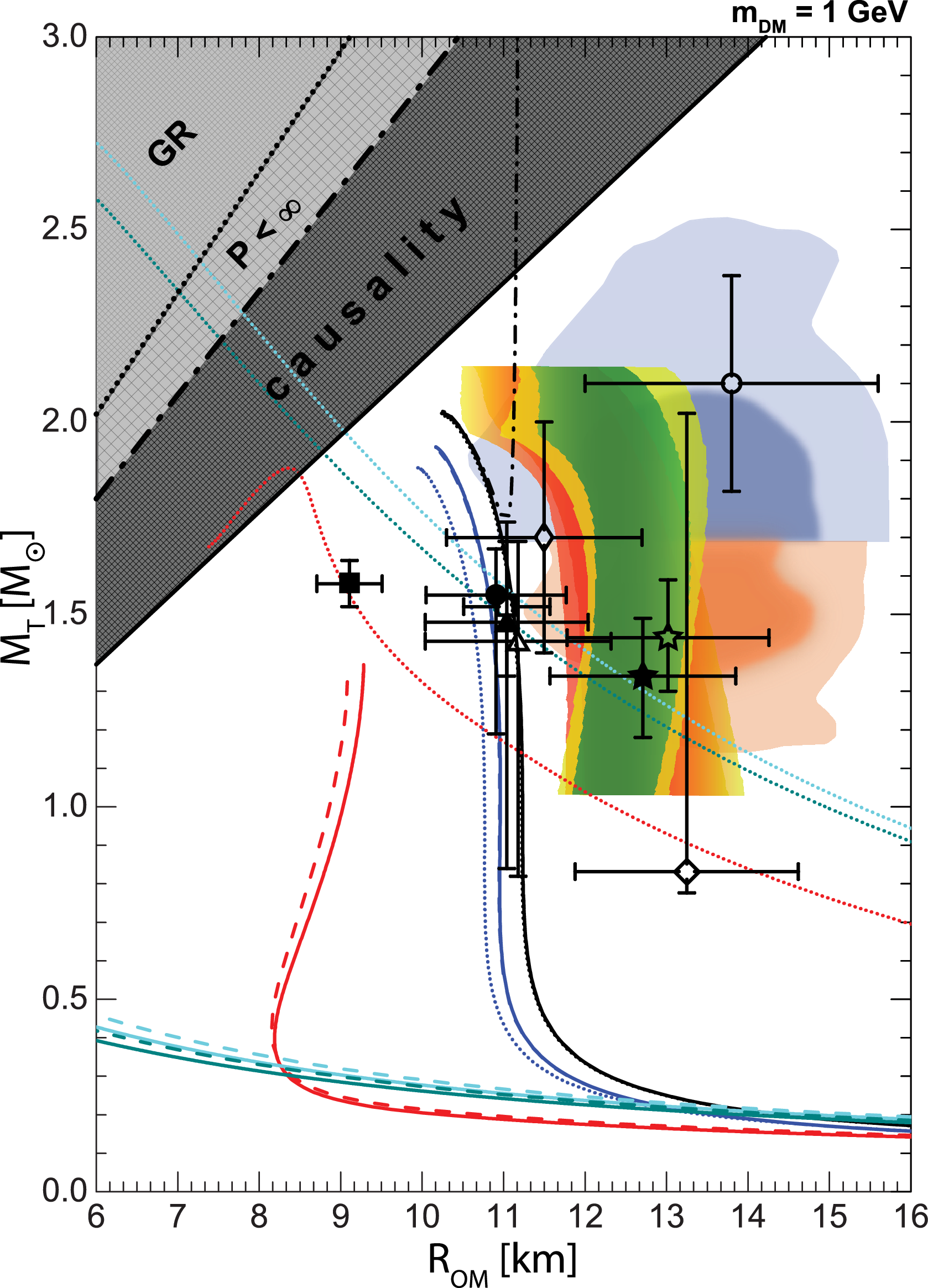}}
\subfigure[]{\includegraphics[scale=0.20]{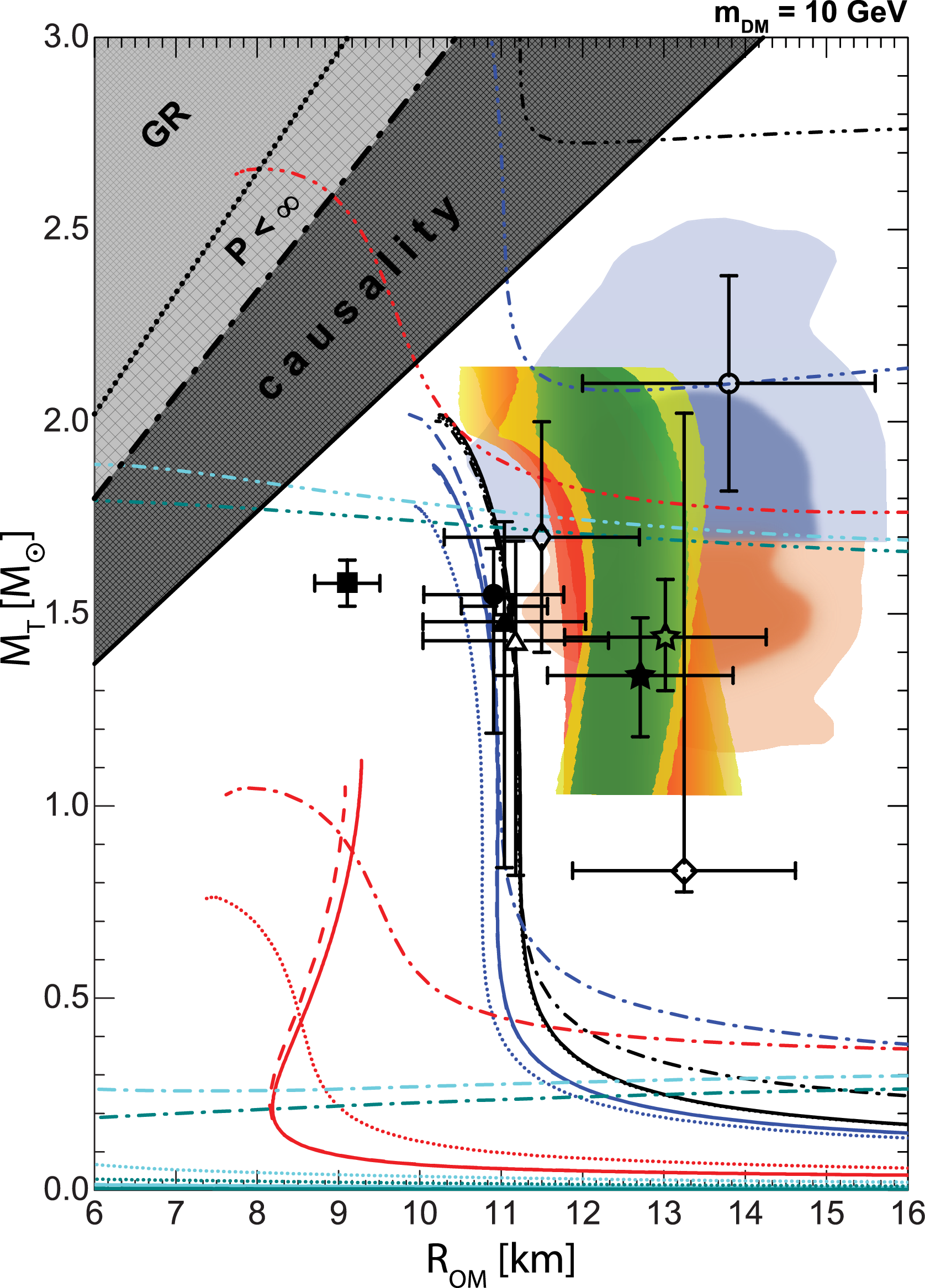}}
\subfigure[]{\includegraphics[scale=0.20]{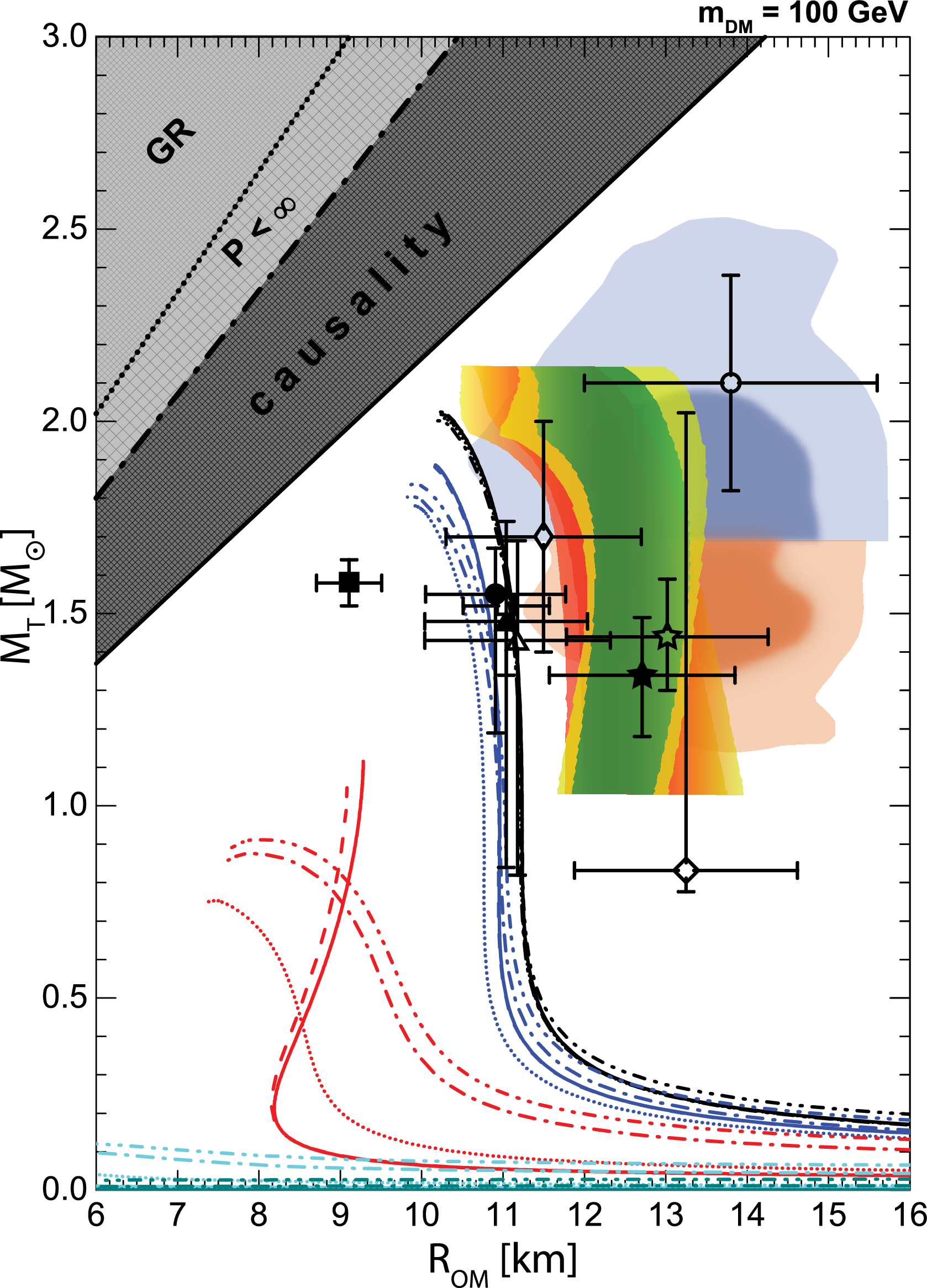}}	
%\end{minipage}\hspace*{-1.8cm}
\caption{
Effects of DM cores on the mass-radius relation for NSs. 
There are five families of lines, each correspond to the different DM interaction strength parameter, $y=0.1;1;10;10^{2};10^{3}$. 
Different line type denotes the $M-R$ relations coming from the different $p^\prime_{DM}/p^\prime_{OM}$ assuming DM particle mass of 1-100 GeV \citep{Tolos:2015qra, Deliyergiyev:2019vti}. 
Data points are the experimentally observed NSs. 
Diagonal lines in the top left corner denote the regions excluded by GR. The colored regions denote the probability distribution functions with the combined constraints on maximum mass and tidal deformability 
computed using signal from the inspiral  of a binary NS system, event GW170817 \citep{Most:2018hfd}. The red and orange bands indicate the 2$\sigma$ and 3$\sigma$ confidence levels, respectively. The green and yellow bands indicate the tune of the 2$\sigma$ and 3$\sigma$ confidence levels by consideration neutron matter in the outer core using prescription of Ref.\citep{Drischler:2017wtt}. The 50\% (light blue (light brown)) and 90\% (deep blue (deep brown)) credible limits for the component masses ($m_{1}$ in blue, $m_{2}$ in brown) and radii computed using signal from the inspiral of a binary NS system, event GW190425 \citep{Abbott:2020uma}.
}
\label{fig:M-R_mDM1_100}
\end{figure*}
%=====================================	

Since, we do not know the exact EoS for DM, therefore one needs to think about a fermionic or bosonic gas with or without interactions. 

We applied a repulsive interaction for the DM fermion gas, so that an increase in the number density increases the pressure and energy density. Therefore, following prescription of Ref.\citep{Narain:2006kx} and using the same notations 
the DM EoS can be rendered like:
\begin{align}
	\label{eq:DM_EoS_presure}
	\rho^{\prime} = &\frac{1}{8\pi^{2}} \left[ (2z^{3}+z)(1+z^{2})^{\frac{1}{2}} - {\rm{\sinh}}^{-1}(z)   \right] + \left(\frac{1}{3\pi^{2}}\right)^{2} y^{2}z^{6}\\ \nonumber
	p^{\prime} = &\frac{1}{24\pi^{2}} \left[ (2z^{3}-3z)(1+z^{2})^{\frac{1}{2}} + 3{\rm{\sinh}}^{-1}(z) \right] \\ 
	&+\left(\frac{1}{3\pi^{2}}\right)^{2} y^{2}z^{6}\nonumber
\end{align}
where $z=k_{F}/m_{f}$ is the dimensionless Fermi momentum. 
The interaction strength, $y=m_f/m_I$, is expressed in terms of the ratio of the fermion mass, $m_f$, to the energy scale of interaction $m_I$.	
For each value of $y$ there is a different EoS and two different regimes exist, namely the non-relativistic limit with $z\ll 1$, and the relativistic limit with $z\gg 1$. For small $y\ll 1$, the EoS will be similar to an ideal Fermi gas while for large $y\gg 1$, the EoS will be mostly determined by the interaction term, unless $z$ becomes small enough so that the EoS becomes dominated by the ideal gas term. 

The repulsive vector interactions considered in that EoS are 
applied in our paper.  More sophisticated and realistic interaction models, which are used for baryonic matter \citep{Serot:1984ey} as well as for quark stars \citep{Hanauske:2001nc}, should also include an attractive scalar interaction. 
Such 
is not the case in the current paper.

In this paper, we examined the wide range of DM interaction strengths 
with the two extreme values, namely a weakly interacting DM ($m_I \sim 100~{\rm MeV}$) with $y=10^{-1}$, and strongly interacting DM ($m_I \sim 300~{\rm GeV}$) with $y=10^{3}$. The three intermediate values of $y$, are 1, 10, and 100. 
In Ref.~\citep{LI:2013qqa} author uses the opposite name convention, namely for strongly interacting DM particles, $m_I \sim 100~{\rm MeV}$, according to the gauge theory of the strong interactions; for weakly  interacting DM particles, $m_I \sim 300~{\rm GeV}$, which can be interpreted as the expected masses of $W$ or $Z$ bosons generated by the Higgs field.

The considered range of $y$ covers almost a full region of interests, since small values of $y$ the mass-radius relation remains almost unchanged as the kinetic terms dominate the EoS, \eqref{eq:DM_EoS_presure}. Any change in the equation of state can only occur for $y \geqslant 1$, i.e. when the interaction term starts dominating the EoS, before the fermions become relativistic. It is thus sufficient to take $y$ in the range $10^{-1}$ to $10^{3}$. Below $y=10^{-2}$, hardly any change in the mass-radius relation can be observed \citep{Narain:2006kx}.

Similarly to Ref.~\citep{Deliyergiyev:2019vti}, and in order to handle the solution space in a more controlled way, the dimensionless ratio between the DM pressure and that of the ordinary matter, $p^{\prime}_{DM}/p^{\prime}_{OM}$, is assumed to be constant throughout the star, and is restricted in the range of $10^{-5}$ to $10^{5}$.
	
%%%%%%%%%%%%%%%%%%%%%%%%%%%%%%%%%%%%%%%%%%%%%%%%%%%%%%
%%%%%%%%%%%%%%%%%%%%%%%%%%%%%%%%%%%%%%%%%%%%%%%%%%%%%%
\section{Results}
\label{sec:Results} 

In this section, we 
discuss the results of the TOV integration, how the admixture of DM to OM modifies the NSs structure, and what this implies for the hyperon puzzle.

Using the total charge conservation theorem \citep{Norman2000:sv} one 
can compute the number of baryon (resp., dark matter particles) in a given star by integrating the divergence of the conserved baryon number current over its invariant volume element, $\sqrt{\rm{det}(g_{\mu\nu})}d^{4}x$. For a static spherical star, we get
\begin{equation} 
	N_{OM, DM}=4\pi \int_{0}^{R} dr r^{2} \rho_{OM, DM}(r) \left[ 1- \frac{2M(r)}{r} \right]^{-1/2}
	\label{eq:NumberOfBaryons}
\end{equation}
with the baryon (respectively, dark matter particles) number density noted 
$\rho_{OM, DM}$. At the surface of the largest star component (DM or baryons) ${\rm{max}}(R_{OM}, R_{DM})$, the outer vacuum results in an outer Schwarzschild solution. In the coordinate radius range of the star, the maximum mass depends on the relative DM to total mass, see Fig.~\ref{fig:NS_MTvsRom}. Therefore, the NS mass-radius relation is clearly affected by even a few percent of DM, with a smaller maximum mass for a larger DM percentage. One can generalise the results from Fig.~\ref{fig:NS_MTvsRom} to $N_{DM} > N_{OM}$, or  $M_{DM} > M_{OM}$, by the switching $DM\leftrightarrow OM$. Thus, admixtured DM NS could appear with a baryonic compact object of only a few kilometers radius.

The relativistic mass-limit considered here gives a qualitative picture of how the critical mass of the core is modified when DM content is present, or  
accreted from the environment \citep{DelPopolo:2019nng}. 
Estimates of the accumulated DM can be made using the Galactic DM profiles \citep{Navarro1997,Stadel2009, Navarro2010, DiCintio2014} combined with the accretion rate estimates given in several papers \citep{Kouvaris2008, Kouvaris:2010vv, deLavallaz:2010wp, Kouvaris:2013awa}. 
Results of our numerical calculation are presented in Fig.\ref{fig:NS_MTvsRom}, showing the NSs mass-radius relations in terms of the percentage of DM acquired, 
\begin{equation}
	{\rm f_{dm}}=M_{\rm DM}/(M_{\rm OM}+M_{\rm DM}), 
\end{equation}
for different particle mass, and different DM interaction strength parameter 
$y$  
\citep[see Ref.~][ for details]{Tolos:2015qra, Deliyergiyev:2019vti}\footnote{The repulsive interaction strength, $y$, among the DM particles is assumed to be a free parameter}. $M_{OM}$ and $M_{DM}$ are calculated by the product of the particle mass and total number of particles for OM and DM in the star. They may be referred to as the baryonic masses for OM and DM (though it should be noted that DM is nonbaryonic).  
In the figure, $M_{\rm{max},T}$ is the gravitational total maximum mass of the star, and $R_{\rm{max}}$ is the radius of the star at the maximum mass. 
When ${\rm f_{dm}}$ 
is close to zero, or when $p^\prime_{DM}/p^\prime_{OM}$ is below $10^{-3}$, the NS can be approximated as an ordinary NS without DM. The 
maximum total masses of these configurations are fluctuating around $2M_{\odot}$.

The top panels of Fig.\ref{fig:NS_MTvsRom} show how the NS radius changes with the DM content at the maximum gravitational total mass. In the case of particles mass 1 GeV, and for $y=0.1$, 1, 10, the NS radius decreases with a larger quantity of DM accumulated. On the  contrary, in the cases, $y=100$, $10^{3}$, the NS radius significantly increases for 
the same fraction of DM, ${\rm f_{dm}}$, and the same $p^\prime_{DM}/p^\prime_{OM}$. The magnitude of such NS radius inflation depends on $y$ and DM particle mass.  
The cases $m_{\rm dm}=10$ and 100 GeV show the NS radius contraction with a larger DM content. 
Moreover, the larger is DM interaction strength, the more steep changes of the NS radius with further increase of $p^\prime_{DM}/p^\prime_{OM}$, at the point when ${\rm f_{dm}}$ 
is close to 100\% the star radius undergoes an abrupt decrease. Thus, we conclude that heavy DM particles tend to create very compact core \citep{Deliyergiyev:2019vti}, which even despite a small fraction, ${\rm f_{dm}}$, reduces $M_{T}$ of NS. 

The bottom panels show how  
$M_{\rm{max},T}$ changes with the DM content. In the case
$m_{\rm dm}=1$ GeV, and for $y=0.1$, 1, the NS mass decrease with increasing content of DM. The case $m_{\rm dm}=100$ GeV shows a mixed behavior: 
a decrease of NS mass with increasing content of DM, till ${\rm f_{dm}}%\epsilon 
\simeq 60 \%$, followed by an increase creating an ankle. We see, that the mass of DM particle has a significant effect on the $M-R$ relations of NSs \citep{Deliyergiyev:2019vti, Ivanytskyi:2019wxd}. 
Such a sensitivity of the NSs mass to the presence of DM is related to its distribution in the stellar interior Fig.\ref{fig:EnerDenstDM_vs_Rom}. 
We see 
there that DM energy density has a nonlinear dependence on  
$R$, and slowly decreases as the NS radius approaches 
limiting values. 

In the cases $y=100$, $10^{3}$, the NS masses may yield $> 2M_{\odot}$ with increasing DM content for light DM particles.
At small, $m_{\rm dm}$, DM does not form a compact structure inside the NS, which may significantly reduce its total mass \citep{Deliyergiyev:2019vti}. 

Moving towards particle mass $m_{\rm dm}=100$ GeV, the total mass decreases with the DM content, with a slower decrease for the weakly interaction strength parameters, $y$. For higher $y$, the decrease of the total masses is steeper, with DM interaction strength parameters differences only discernable at the high end of the DM contribution to the total star mass.
We admit, that the picture that we observe is a bit complicated with respect to linear dependence shown in \citep{Mukhopadhyay:2015xhs}. Qualitatively it is close to observation reported in Ref.\citep{Ivanytskyi:2019wxd}.

We thus conclude that the interplay between the small DM particles mass and the DM interaction strength parameter indeed modifies significantly the mass-radius relation, 
while heavier DM particles lead to a substantial reduction of the NS maximal mass and radii in the full range of the examined $y$ with increase of the DM fraction. 
In addition, it is worth noticing that the changes, at the same level of DM contribution to the NS content, are less pronounced for radius rather than for the total mass.

In order to get more insight into the radius, and maximum total mass dependencies on the particle mass of DM candidates, we show in Fig. \ref{fig:MT_pDMpOM} that, for  
a given proportion of DM inside the stars, 
a higher DM particle mass in general leads to a smaller total mass. 
This can easily be understood by noting that, since the OM and DM are assumed to be non-interacting (except through gravity), the DM core is only supported 
by its own degenerate pressure. It is well known that the maximum mass limit for a self-gravitating Fermi gas decreases as the particle mass increases. We observe an inverse 
logarithmic dependence. At very low values of $p^\prime_{DM}/p^\prime_{OM}$, below $10^{-3}$, this dependence is almost flat, 
which can be noted from the region of very low ${\rm{f_{dm}}} < 10^{-5} \%$ on Fig.\ref{fig:NS_MTvsRom}.

Hence, the onset of the collapse of a degenerate DM core is responsible for the dependence of $M_{\rm{max,T}}$ on $m_{\rm dm} $ as seen in Fig. \ref{fig:MT_pDMpOM}. 
It should also be noted that, while the pressure of OM within the DM core does not contribute to supporting the weight of the DM core, the mass of the OM fluid does enhance the collapse of the DM core. 
The total pressure (energy) density is a simple sum of the DM pressure (energy) and NS pressure (energy). 
Fig. \ref{fig:MT_pDMpOM} also shows that a given maximum total mass of the NS is obtained 
by a combination of $m_{\rm dm}$, $y$, and $p^\prime_{\rm DM}/p^\prime_{\rm OM}$. As an example, in the case $p^\prime_{\rm DM}/p^\prime_{\rm OM}=10^{3}$ a NS with mass of $1M_{\odot}$ can be obtained for $y=0.1$, and $m_{\rm dm} \simeq 1$ GeV. 
In the case $y=10^{3}$ a NS mass of $2M_{\odot}$ can be obtained for $m_{\rm dm} \simeq 10$ GeV. 
A very high NS mass, 
much larger than $2M_{\odot}$, could be achieved when the particle mass is small enough. Such kind of dark-matter-admixed NSs could explain the recent measurement of the Shapiro delay in the radio pulsar PSR J1614-2230 \citep{Demorest:2010bx}, that may hardly be reached if hyperons alone are considered, as in the case of the microscopic Brueckner theory \citep{Burgio:2010ek, Burgio:2011wt}. We conclude that a very high mass measurement around 
$2M_{\odot}$ requires a really stiff EoS in NSs. 

Fig. \ref{fig:MTRom_m1_500_y01_103} plots the mass-radius relation for the different $p^\prime_{\rm DM}/p^\prime_{\rm OM}$, $y$, and $m_{\rm dm}$. 
It gives complementary information to Fig. \ref{fig:MT_pDMpOM} for the same range of $p^\prime_{\rm DM}/p^\prime_{\rm OM}$. 
The plot shows several trends: the NS total mass: 
\begin{enumerate}
	\item decreases with DM particle mass; 
	\item increases with increasing interaction strength, $y$; 
	\item different values of $p^\prime_{\rm DM}/p^\prime_{\rm OM}$ mainly influence the NS total mass for large values of the DM particle mass.
\end{enumerate}  

In Fig. \ref{fig:M-R_mDM1_100}, we compared predictions of our model with the observational data. However, we only considered 
NS reports which provided at least a pair of 
observational parameters such as mass and radius. Our list included the following objects: 
EXO 0748-676, EXO 1745-248, 4U 1608-522, 4U 1820-30, M13, wCen, X7, RX J1856-3754, EXO1722-363. In addition, we considered the most recent NS observation, PSR J0030+0451 \citep{Miller:2019cac, Riley:2019yda} and NS-NS merger, event GW170817 \citep{Most:2018hfd}. 
We plotted the $M-R$ relation for several values of the ratio $p^\prime_{\rm DM}/p^\prime_{\rm OM}$, interaction strength, $y$, and DM particle mass, $m_{\rm dm}$. 
In order to keep the figure 
readable, we did not plot the lines corresponding to different percentage content of DM, ${\rm{f_{dm}}}$. By the colored regions we denoted the probability distribution functions with the combined constraints on maximum mass and tidal deformability computed using signal from the inspiral of a binary NS system, event GW170817 \citep{Most:2018hfd}. 
The red and orange bands indicate the 2$\sigma$ and 3$\sigma$ confidence levels, respectively. The green and yellow bands indicate the tune of the 2$\sigma$ and 3$\sigma$ confidence levels by consideration neutron matter in the outer core using prescription of Ref.\citep{Drischler:2017wtt}. 
There are dramatic differences among various 
DM particle mass and DM content with the 
same EoS. The presence of a maximum mass for each $p^{\prime}_{\rm DM}/p^{\prime}_{\rm OM}$ is clearly apparent.

As we already reported, observational results relative to masses and radii of some NSs (e.g., Vela X-1, 4U 1822-371, PSR J1614-2230, PSR J0348+0432) contradict theoretical predictions of ``normal'' NSs. Ciarcelluti, in \citep{Ciarcelluti:2010ji}, showed that a DM core in NSs can explain the unusual properties of those NSs. 
For instance, the NS, such as EXO 0748-676 can be explained with $m_{\rm dm}= 10$ GeV, $p^{\prime}_{\rm DM}/p^{\prime}_{\rm OM}=10^{-1}$, and $y=10^{3}$. 
However, as shown by 
Ref.\citep{Ciarcelluti:2010ji}, EXO 0748-676 can be explained with models without DM. 
The very light weight NSs, like X7, can be reproduced with $p^{\prime}_{\rm DM}/p^{\prime}_{\rm OM}=1$ for $y=1$. 
With use of the EoS from H4 model \citep{Glendenning:1991es}, one may also provide a good description of such star masses observations.

The values of the $p^\prime_{\rm DM}/p^\prime_{\rm OM}$ in the range of $0.01-0.1$ may explain observational $M-R$ relations for EXO 1745-248, 4U 1608-522, M13, wCen, RX J1856-3754 if one 
assumes the range of DM interaction strength 
$0.1 - 1$. 
The strongly interacting cases are only acceptable for light DM particles, for 
$y$ in the range $0.1-10$, and all the $m_{\rm dm}$ considered in Fig.\ref{fig:M-R_mDM1_100}. 

A distinctive property of the DM dominated stars is their small OM core radius, 
about a few km, from which thermal radiation could be observed. The detection of a compact star with a thermally radiating surface of such a small size could provide strong evidence for their existence. 
In this view, the most compact NS, 4U 1820-30, 
can be reproduced with $p^{\prime}_{\rm DM}/p^{\prime}_{\rm OM}=1$, $y=10$, assuming $m_{\rm dm}=1$ GeV.

From 
Fig. \ref{fig:M-R_mDM1_100} one may note many 
intersection points between different $M-R$ curves. While the DM stars have the same $M$ and $R$, their internal structures in fact differ significantly. For instance, they 
will have different DM to OM pressure ratios. 
Since $M$ and $R$ of the stars at these intersection points are same, it would seem impossible to distinguish them based on their gravitational effects on other nearby stellar objects.

%%%%%%%%%%%%%%%%%%%%%%%%%%%%%%%%%%%%%%%%%%%%%%%%%%%%%%
%%%%%%%%%%%%%%%%%%%%%%%%%%%%%%%%%%%%%%%%%%%%%%%%%%%%%%	
\section{Constraints on Limiting Mass}
\label{sec:Constraints}

The critical mass of the core of ordinary stars is modified in the presence of the DM particles, because they 
act as an additional source of gravity. Such stars therefore collapse at lower core masses and this could result in extraordinary compact NS with low mass, due to the presence of DM.

We are interested in cores with high density that are near gravitational collapse. The densities, essentially given by the rest-mass density of nucleons, $\rho_{0}$, and the dominant contribution to the pressure, $p$, come from ultrarelativistic electrons. 

NSs are mainly made of strongly degenerate neutrons.  Nevertheless, inverse $\beta$-decay equilibrium allows  protons, electrons and muons to also be present,  though in lower fractions \citep{Gandolfi:2019zpj}.
The exact microscopic composition of NSs depends on the number density radial profile. 
To precisely determine the internal structure of a NS 
the EoS of dense matter is a key ingredient.

\begin{figure*}%[h]
	\centering
	\includegraphics[width=1\linewidth%scale=0.60
	]
	{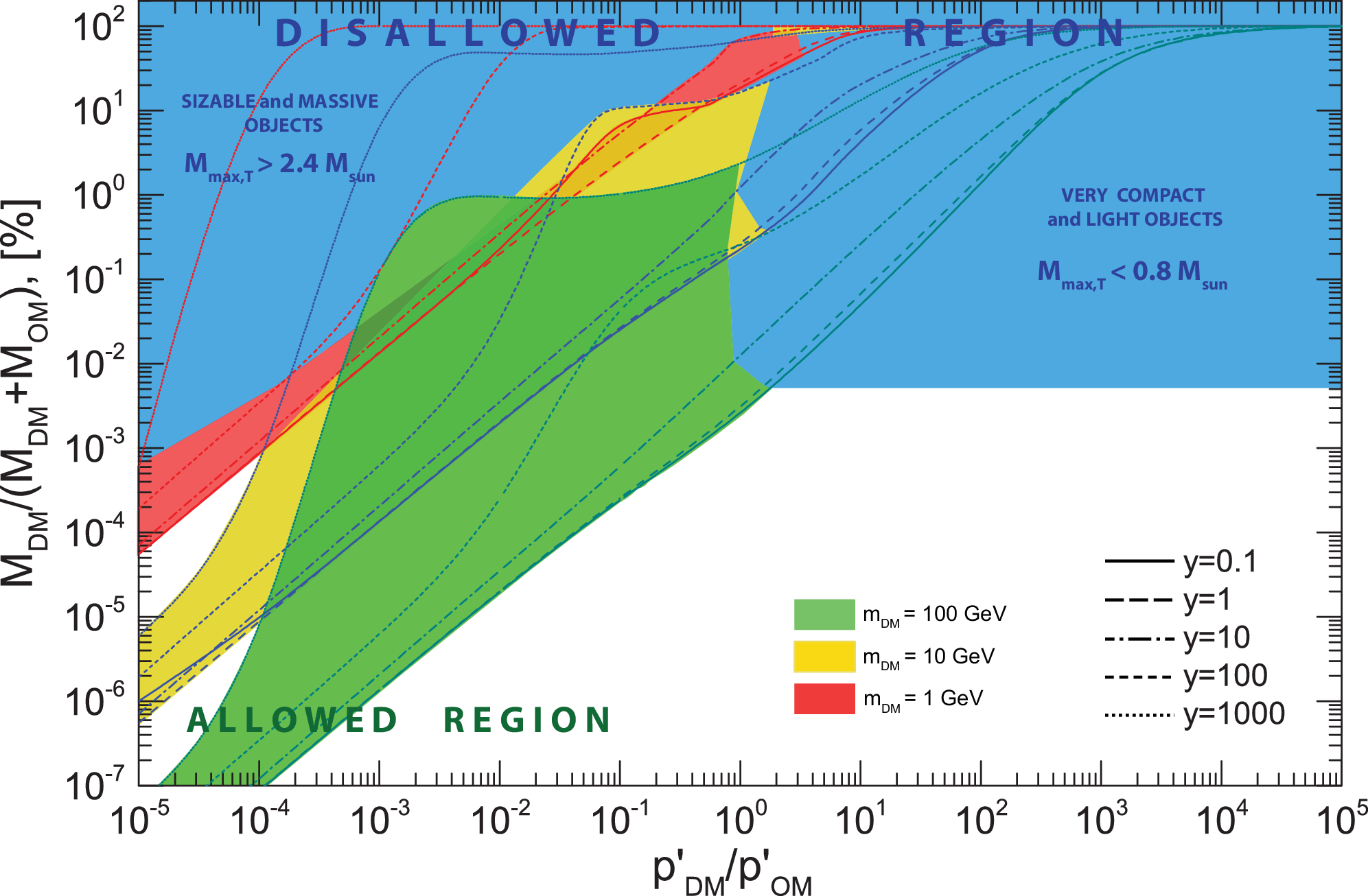}		
	\caption{
		The critical fraction of DM vs. the dimensionless pressure ratios, $p^\prime_{\rm DM}/p^\prime_{\rm OM}$. 
		The blue area in the top left corner represents the unphysical region with very sizable and massive NSs, $M_{\rm{max,T}} > 2.4M_{\odot}$.  While the blue area on the right represents the unphysical region with very compact and light objects  $M_{\rm{max,T}} < 0.8M_{\odot}$. 
		The red, yellow and green bands are the allowed regions for NSs with admixtured DM for $m_{\rm {dm}}=1, 10, 100$ GeV respectively, namely the regions which fall into $0.8-2.4 M_{\odot}$ mass range for NSs. The curved lines denote DM fraction as function of $p^\prime_{\rm DM}/p^\prime_{\rm OM}$ for the given DM particle mass and were added as a guide.
	}
	\label{fig:ADM_NS_constraints}
\end{figure*}

All of the EoSs proposed so far for the description of the complexity of the interactions and the structure of NSs use certain approximations and involve numbers of fundamental parameters. One is usually 
checking some of the proposed EoSs against the data collected 
in 
experiments, however another possible approach 
is to parameterize plausible 
EoSs with a few phenomenological parameters and to analyse which constraints are imposed by observations on these parameters. With this in
mind, we  
tried 
to put constraints on the physical region defined by 
the critical fraction of DM versus the dimensionless pressure ratios, $p^\prime_{\rm DM}/p^\prime_{\rm OM}$, assuming the loose range $0.8-2.4 M_{\odot}$ for NSs masses 
\citep{Chamel:2013efa, Ozel:2012ax}. This introduces a direct uncertainty on the relative prediction of the contrained regions. This approach is applied to evaluate the impact of changes to the $M_{{\rm max},T}$ as a function of DM fraction, sensitivity to the DM interaction strength of the model. 
The results shown on Fig.\ref{fig:NS_MTvsRom} enabled us to map critical values of ${\rm{f_{dm}}}$ into the $p^{\prime}_{\rm DM}/p^{\prime}_{\rm OM}$ axis. 
The dependence of DM fraction, ${\rm{f_{dm}}}$, on DM particle mass at 
the same time helps us to constraint the DM particle mass from Fig.\ref{fig:ADM_NS_constraints}. The broad range of the examined $y$ parameters provides a more general view on the constraints regions. 
Each of the allowed regions for the given DM particle mass has been formed by the DM fraction profiles corresponding to $y=0.1-1000$, which are denoted by the solid, dashed, dash-dotted, short dashed and dotted curves respectively. 
We see that, 
moving towards 
low DM particle masses, the allowed regions start to get squeezed. Regions with $p^{\prime}_{\rm DM}/p^{\prime}_{\rm OM} > 2$ are disallowed as they produce 
very compact and light NSs, while regions with $p^{\prime}_{\rm DM}/p^{\prime}_{\rm OM} < 2$ and ${\rm f_{dm}} > 0.01\%$ are disallowed as they output too 
sizable and massive NSs.

We point out that the choice of parameters in Fig.~\ref{fig:ADM_NS_constraints} is not always allowing discrimination of different DM masses and interaction strength, $y$, since 
there are overlaps between corresponding allowed regions. At the current stage, astrophysical observation for NS can be sensitive to the low mass DM particle, in the case of the weakly interacting DM when $y \leqslant 10$ (see dotted and short dashed curves on 
Fig.~\ref{fig:ADM_NS_constraints}). However, if DM particle mass is greater than 1 GeV, one may expect that DM interaction strength  is stronger, while the value of $p^{\prime}_{DM}/p^{\prime}_{OM}$, should be less than $\sim 3$. Using the constraints presented on Fig.~\ref{fig:ADM_NS_constraints}, the search for DM effects in NSs mass range ($0.8-2.4 M_\odot$) can get more complicated. A more robust way to discriminate DM effects uses the 
search for unconventional NS masses, reported in Ref.\citep{Narain:2006kx, Tolos:2015qra, Deliyergiyev:2019vti, DelPopolo:2019nng}. In this case, with correspondingly enlarged allowed regions,  the constraints shown on Fig.~\ref{fig:ADM_NS_constraints} can also be relevant.
Furthermore, adding the stringent DM particle mass limits from the direct detection experiments on top of these regions would require 
making some assumptions. 

Each of the allowed regions for the given DM particle mass has been formed by the DM fraction profiles corresponding to $y=0.1-1000$, which are denoted by the solid, dashed, dash-dotted, short dashed and dotted curves respectively. 
We see that, 
moving towards 
low DM particle masses, the allowed regions start to get squeezed. Regions with $p^{\prime}_{\rm DM}/p^{\prime}_{\rm OM} > 2$ are disallowed as they produce 
very compact and light NSs, while regions with $p^\prime_{\rm DM}/p^\prime_{\rm OM} < 2$ and ${\rm f_{dm}} > 0.01\%$ are disallowed as they output too  
sizable and massive NSs.

%=====================================
%=====================================

%%%%%%%%%%%%%%%%%%%%%%%%%%%%%%%%%%%%%%%%%%%%%%%%%%%%%%
%%%%%%%%%%%%%%%%%%%%%%%%%%%%%%%%%%%%%%%%%%%%%%%%%%%%%%
\section{Conclusions}
\label{sec:Conclusions}

In this paper, we have studied how DM, non-self-annihilating, self-interacting fermionic dark matter, admixed with ordinary matter in NSs changes their inner structure, and discussed the mass-radius relations of such NSs. 
We considered DM particle masses of 1, 10, and 100 GeV, while taking into account a rich list of the DM interacting strengths, $y$. 
While recent studies show that the presence of hyperons in NSs is of crucial importance and seems to be unavoidable, the recent studies with inclusion of the hyperon many-body forces \cite{Vidana:2010ip, Nishizaki:2002ih} 
do not give us a full solution to hyperon puzzle. 
We argue, that including DM in 
NSs hydrostatic equilibrium, namely adding DM into the balance between the gravitational force and the internal pressure may explain the observed discrepancies, regardless of hyperon multi-body interactions.

The total mass of the NSs depends on the DM interacting strength, on DM particle mass as well as the quantity of DM in its interior, DM fraction. By analyzing this multidimensional parameter space we put constraints in the parameter space ${\rm f_{dm}} - p^{\prime}_{\rm DM}/p^{\prime}_{\rm OM}$. 

The DM fraction in a NS may not be the same in all stars, since it may depend on the NS age \cite{Kouvaris2008, Bertone2008, Kouvaris:2010vv, deLavallaz:2010wp}, initial temperature, or on the environment in which it was formed \cite{DelPopolo:2019nng}. Such possibilities 
further complicate the interpretations of the measurements of NS properties.

We note that currently available experimental data, with a present level of uncertainties, falls into the range $y=0.1-10$ in the whole DM particle mass spectra considered in this paper. As was already mentioned, the choise of the EoS 
also may play a crucial role in the data interpretation. Therefore, one has to be 
quite delicate in drawing further conclusions.

On November 1st the LIGO-Virgo Collaborations resumed their search for gravitational waves, which will give us new datasets on NS-NS mergers, and 
will definitively tell us more about the EoS of compact stars, including, perhaps, about possible phase transitions at supranuclear density, and possible DM core effects\citep{Falcke:2013xpa, Fuller:2014rza, Ellis:2017jgp, Ellis:2018bkr}.

Observations 
by the NICER (Neutron star Interior Composition Explorer) mission \citep{NICER:2012} 
have already started 
to provide us 
new 
insights about NSs \cite{Miller:2019cac, Riley:2019yda}. The 
planned eXTP (enhanced X-ray Timing and Polarimetry) Mission \citep{Watts:2018iom}, LOFT (Large Observatory For X-ray Timing) satellite \cite{Wilson_Hodge_2016}, and ATHENA (Advanced Telescope for High Energy Astrophysics) \citep{Barcons:2012zb, Motch:2013wfn, Athena:2014cdf} are seen to have completely different systematic errors, and thus, by combining many measurements with such differing systematics, one can hope to significantly narrow the mass-radius range for NSs.

Using our model of non-self annihilating, self-interacting DM particles, we were able to provide predictions on the masses and pressure ratios of viable and stable NS produced by OM admixed with such DM (see Fig.~\ref{fig:ADM_NS_constraints}).

\acknow{M.D. work was supported by the Polish National Science Centre (NCN) grant 2016/23/B/ ST2 / 00692. MLeD acknowledges the financial support by Lanzhou University starting fund and the Fundamental Research Funds for the Central Universities (Grant No.lzujbky-2019-25). %L.T. acknowledges support from the FPA2016-81114-P Grant from Ministerio de Ciencia, Innovacion y Universidades, Heisenberg Programme of the Deutsche Forschungsgemeinschaft under the Project Nr. 383452331 and PHAROS COST Action CA16214.
}
\showacknow % Display the acknowledgements section

%%%%%%%%%%%%%%%%%%%%%%%%%%%%%%%%%%%%%%%%%%%%%%%%%%%%%%
%%%%%%%%%%%%%%%%%%%%%%%%%%%%%%%%%%%%%%%%%%%%%%%%%%%%%%
\appendix
\section{inner/outer  crust EoSs}
\label{sec:Appendix01}

Following the prescription in Ref.\cite{Negele:1971vb,Ruester:2005fm} the EoSs for the inner and outer crust are defined as 	%----------------------------------------------------
\begin{align}
	{\rm{inner~crust}}: &P=n_{b}^{2}\frac{\partial E_{\rm{tot}}}{\partial n_{b}}, ~\Gamma=\frac{n_{b}}{P}\frac{\partial P}{\partial n_{b}},~ \rho=\frac{n_{b}E_{\rm{tot}}}{c^{2}}.\\
	{\rm{outer~crust:}} &P=P_{e}+\frac{1}{3}W_{L}n_{N},~ \Gamma=\frac{n_{b}}{P}\frac{\partial P}{\partial n_{b}},~ \rho=\frac{E_{\rm{tot}}}{c^{2}},
	\label{eq:innerCrust_EoS}
\end{align}
%----------------------------------------------------
where $P$ is the pressure, $\Gamma$ is the adiabatic index, and $\rho$ is the mass density. $P_{e}=\frac{1}{3\pi^{2}}\int_{0}^{k_{e}}\frac{k^{4}}{E_{e}}dk$ is the electron pressure, where $k_{e}$ is the electron Fermi momentum, and $E_{e}=\sqrt{k^{2}+m_{e}^{2}}$. 
$n_{b}=An_{N}$ is the baryon density in baryons per cm$^{3}$, $E_{\rm{tot}}$ is the total energy per baryon in MeV that's defined as
%----------------------------------------------------
\begin{align}
	{\rm{inner~crust:}} E_{\rm{tot}}=m_{n}+c_{0}+{\rm{exp}}\left(\sum_{I=1}^{7}c_{I}x^{I-1}\right),
	\label{eq:innerCrust_Etot}
\end{align}
$x={\rm{ln}}(n_{b}\times 10^{-35})$. $c_{0},c_{I}$ are found from the fits of \eqref{eq:innerCrust_Etot}, separately for the 
ground state and for the uniform neutron gas configurations, see Table 4 in Ref.\cite{Negele:1971vb}. While for outer crust the total energy is defined as:
\begin{align}
	{\rm{outer~crust:~}} E_{\rm{tot}}=n_{N}(W_{N}+W_{L})+E_{e}.
	\label{eq:outerCrust_Etot}
\end{align}
The body-centred 
cubic (bcc)\footnote{used to arrange nuclei in Ref.\cite{Ruester:2005fm}} lattice energy, $W_{L}$, and the energy of the nuclei, $W_{N}$, is defined as 
%----------------------------------------------------
\begin{align}
	W_{L}&=-1.81962\frac{Z^{2}e^{2}}{4\pi\epsilon_{0} a}\\
	W_{N}&=m_{n}(A-Z)+m_{p}-bA,
	\label{eq:outerCrust_W}
\end{align}
%----------------------------------------------------
where $m_{n}$, $m_{p}$ is the neutron and proton mass respectively, $\epsilon_{0}$ is the absolute vacuum permittivity.

% Bibliography
%\bibliographystyle{apsrev4-1}
\bibliographystyle{pnas}	
%\bibliography{DMStarsClose2GC}

\end{document}